\documentclass[prb,twocolumn,showpacs,preprintnumbers,amsmath,amssymb]{revtex4}

\usepackage{amsfonts}
\usepackage{latexsym}
\usepackage{graphicx}

\newcommand{\Eqref}[1]{Eq.~(\ref{#1})}

\newcommand{\nn}{\nonumber}
\newcommand{\be}{\begin{equation}}
\newcommand{\ee}{\end{equation}}
\newcommand{\alf}{{Alfv\'en~}}
\newcommand{\bear}{\begin{eqnarray}}
\newcommand{\ear}{\end{eqnarray}}

\begin{document}

\title{On the theory of MHD waves in a shear flow of a magnetized
turbulent plasma}

\author{T.M.~Mishonov}
\email[E-mail: ]{mishonov@phys.uni-sofia.bg}
\author{Y.G.~Maneva}
\email[E-mail: ]{yanamaneva@gmail.com}
\author{Z.D.~Dimitrov}
\email[E-mail: ]{Zlatan.Dimitrov@gmail.com}
\author{T.S.~Hristov}
%
\affiliation{Department of Theoretical Physics, Faculty of Physics,\\
University of Sofia ``St.~Clement of Ohrid'',\\
5 J. Bourchier Blvd, BG-1164 Sofia, Bulgaria}

\date{\today}

\begin{abstract}
The set of equations for magnetohydrodynamic (MHD) waves in a shear
flow is consecutively derived. The proposed scenario involves the
presence of a self-sustained turbulence and magnetic field. In the
framework of Langevin--Burgers approach the influence of the
turbulence is described by an additional external random force in
the MHD system. Kinetic equation for the spectral density of the
slow magnetosonic (Alfv\'enic) mode is derived in the short
wavelength (WKB) approximation. The results show a pressing need for
conduction of numerical Monte Carlo (MC) simulations with a random
driver to take into account the influence of the long wavelength
modes and to give a more precise analytical assessment of the short
ones. Realistic MC calculations for the heating rate and shear
stress tensor should give an answer to the perplexing problem for
the missing viscosity in accretion disks and reveal why the quasars
are the most powerful sources of light in the universe. The planned
MC calculations can be incorporated in global models for accretion
disks and also in all other physical conditions where there is a
shear flow in a magnetized turbulent plasma. It is supposed that the
heating mechanism by \alf waves absorption is common for many kinds
of space plasmas from solar corona to active galactic nuclei and the
solution of these longstanding puzzles deserves active
interdisciplinary research. The work is illustrated by numerical
calculations and by exact solutions for the time dependence of the
magnetic field given by the Heun function.
\end{abstract}



\maketitle
\section{Introduction}
 The purpose of the present work is to give a detailed derivation of
the stochastic magnetohydrodynamic (MHD) set of equations for a
shear flow in a magnetized turbulent plasma. The meticulously
performed analysis of the MHD system sets the basis for further
Monte Carlo (MC) calculations devoted to reveal the basic physical
phenomena in accretion disk plasmas: their heating and the origin of
a large effective viscosity significantly exceeding the bare plasma
one.
 The paper is organized as follows: short historical remarks
introducing to the readers the motivation for the current work are
presented in Sec.~\ref{sec:Motivi}. Basic wave kinematics in a shear
flow with a transition from Eulerian to Lagrangian coordinates is
given in Sec.~\ref{sec:Kinematics}. Derivation and linearization of
the full stochastic MHD set of equations with external random
forcing by turbulence is described in Sec.~\ref{sec:stochMHD}. Plane
waves anzatz in a shear flow is included for separation of variables
and subsequent reduction from partial to ordinary differential
equations is performed in Sec.~\ref{sec:SepOfVar}. The illustrative
2D case is analyzed in Sec.~\ref{sec:2Dwaves}. Secular equation for
the \alf waves amplitudes is solved and the corresponding damping
rate is obtained. Auxiliary problem for the period averaged energy
of an effective oscillator under a white noise is considered. Short
wavelength (WKB) approximation is applied to the \alf waves
amplitude in Sec.~\ref{sec:WKB} and is it shown that the \alf
spectral density obeys an effective Boltzmann equation.
Sec.~\ref{sec:WhiteNoise} treats the Langevin-Burgers MHD, modeling
the influence of the turbulence as a random external force in the
momentum equation. Further on, speculations on the origin of a large
effective viscosity are proposed. Conclusive remarks and future
perspectives are briefly discussed in Sec.~\ref{sec:concl}. It is
debated on the kind of numerical analysis which has to be done in
order to reveal the origin of the huge effective viscosity, observed
in the accreting magnetized turbulent plasma.

All the analytical calculations utilized in the current work, as
well as some general concepts for heating, related to the
stochasticity of the investigated turbulent system, are laid out in
five separate appendices. Matrix presentation for Lagrange--Euler
transformations in the presence of a shear flow is reproduced in
appendix \ref{app:matrixpresent}. Linearization of the dynamic
equations is performed in Appendix \ref{app:dyneqs}. Detailed
derivation of the complete MHD set of equations for a magnetized
plasma in a shear flow under the influence of a random noise is
provided in appendix \ref{app:MHDsyst}. Test examples of MHD waves
with a restricted wave-vector orientation $K_y =0,$  short
wavelengths and small attenuation are presented. The shear rate
dependent damping of Alfv\'en waves propagating along the magnetic
field lines is calculated in appendix \ref{app:koefnazatihv}.
Several illustrative examples for the power rate due to stochastic
heating of a Brownian particle, oscillator under a white noise and a
free particle are applied in \ref{app:stochHeating} to support the
general consideration for a white-noise driven heating, introduced
in sec.~\ref{sec:WhiteNoise}.

\section{Motivation}
\label{sec:Motivi}
%
Matter accreting onto a compact object redistributes radially its
angular momentum, dissipates gravitational energy and forms a
powerful source of radiation through physical processes that still
cannot be regarded as understood. Accreting plasma's temperature and
pressure predict a molecular viscosity that is orders of magnitude
too small to account for the observed radiation intensity. Thus the
origin of a large effective viscosity in the magnetized turbulent
plasma of the accretion disks is a significant and yet unsolved
problem. We know how a laser emits light, how the luminescent lamp
works, we know how flash a fire-fly in the summer nights, but we do
not know yet what is the mechanism of glowing of the most luminous
sources of light in the universe -- we plan to reveal this
long-standing problem.

Frictional release of angular momentum is believed to be an
important element in stars formation as well. In our solar system,
for instance, the large mass concentrated into the sun carries only
two percents of the angular momentum,\cite{Gurevich:78} while all
the rest is associated with the planetary motion. Next to the
Galileo assertion \textit{``Eppur si muove''} (and yet it does
move), for half a century we still face the question ``why does the
Sun not move (rotate) faster''. Presumably, at an early stage of the
system's formation the proto-planetary disk has acted as a brake
which slowed down the rotation of most of the disk's mass, so it
could cluster and ``ignite'' as a star. Broadly, such a mechanism of
angular momentum release appears responsible for the formation of
compact astrophysical objects and possibly determines the universe
in the way that we observe it today. In this context, there still
exists a central open question of what physical process produce the
friction forces that facilitate the stars formation.

The purpose of the present work is to demonstrate a strong energy
dissipation in a shear flow of a magnetized plasma based on a simple
model. Despite the process's interdisciplinary nature, we will
pursue a description that is built on first-principles physics. Our
leading hypothesis is that in a shear flow of the almost inviscid
plasma the Alfv\'en waves get amplified\cite{Chagelishvili:93} and
later have their energy thermalized, i.e. ``lasing'' of Alfv\'en
waves is the basis of dissipation in accretion disks. We select an
approach which is statistical rather than fluid-mechanical, by
employing a kinetic equation for the spectral density (proportional
to the square of their amplitude) of the Alfv\'en waves. In the
spirit of the quantum mechanics the square of the amplitude is
proportional to the number of particles, which we will call
``alfvons''. Here the kinetic equation describes the dynamics of the
alfvons population. The lasing of the media is analogous to the
dynamics of ecological systems where a fast population growth is
followed by a resettlement and high death-rate. Similarly here, the
energy transferred from the shear flow first increases the alfvons
density and is later dissipated, thus effectively raising the
plasma's viscosity and resistivity. Our goal is to deduce a kinetic
equation for the spectral density of the Alfv\'en waves, as well as
its solution, and analyze the physical factors driving the process.
Standard methods of quantum mechanics (quasi-classical approximation
for short wave-lengths and perturbation theory to account for the
small initial viscosity of the plasma) will be considered as
adequate. As an initial step, we employ a simplified treatment of
the turbulence in accretion disks -- the Burgers approach and show
that the turbulence's role in a model approximation may be reduced
to the source term in the kinetic equation. We also assume that it
is the turbulence that triggers the Alfv\'en waves, which are later
amplified by the shear flow. There are many examples where waves can
be amplified but often conditionally one can mention lasers and in
general case lasing processes in some unstable medium.

The considered model problem involves convective instability and
turbulence in the heated disk's plasma and describes a
self-sustained \textit{scenario} for heating in accretion disks. The
velocity fluctuations serve as a random force that initiates the
Alfv\'en waves, which are then subject to a large amplification. The
energy carried by these alfvons is then transformed into heat
through the molecular viscosity and Ohmic resistivity, thus also
creating friction forces. As the heat generation is more intense in
the middle of the disk's thickness, the consequent temperature
difference drives the convective instability and the turbulent
convection. Thus the process sustains itself and the gravitational
energy transforms into heat.

\section{Wave kinematics in shear flows}\label{sec:Kinematics}
%
To describe locally the motion of the accreting fluid, we choose the
$z$ axis along the velocity and the $x$ axis in direction of the
velocity's gradient. Thus for the background shear velocity field we
have
\be \label{vel} U^{(0)}_z = Ax, \quad U^{(0)}_x = U^{(0)}_y = 0 .
\ee
The superscript $(0)$ refers to an equilibrium laminar shear flow
whose perturbations will be studied.

A selected small element of the fluid is carried by the flow so that
its $z$ coordinate is a linear function of time
\be
\label{tagged}
z_{at} (t) = Axt + z_{at} (0).
\ee
Under such drift the coordinates $x,y$  remain unchanged, i.e.
$x_{at} (t) = \mathrm{const},$ $y_{at} (t) = \mathrm{const}$. For a
hydrodynamical description we will use flow-following (Lagrangian)
coordinates
\be \label{newvar}
\tilde{x} = x, \quad \tilde{y} = y, \quad
\tilde{z}= z - Axt.
\ee
These tilde variables $\tilde{\mathbf{r}}$ are the Cartesian
coordinates of the ``tagged'' atoms at the initial moment $z_{at}
(0)=z_{at} (t) - Axt.$ They express the initial position of the
fluid element. The change of the variables (\ref{newvar}), however,
does not alter the projection in the $xy$ plane. As the initial
position of the ``tagged'' atoms is fixed we can consider the tilde
coordinates as ``frozen'' in the fluid.

 Let us now consider a plane wave in the shear flow with an amplitude
$\propto \exp(i\mathbf{k} \cdot \mathbf{r}).$ The requirement of a
phase invariance
\be \label{phase-invariance} \tilde{\mathbf{k}}\! \cdot
\tilde{\mathbf{r}}
 = \mathbf{k}\cdot \mathbf{r}
\ee
sets the transformation law
\be \label{newwavevec}
\tilde{k} _{x} = k_x + Atk_z, \,\,
\tilde{k}_y = k_y, \,\, \tilde{k }_{z} = k_z,
\ee
which can be validated by a substitution of \Eqref{newvar} into
\Eqref{phase-invariance}.

As the initial tilde coordinates are related to the frozen initial
position of the atoms, the wave vector in tilde coordinates is also
``frozen'', $\tilde{\mathbf{k}}=\mathrm{const.}$ This determines the
evolution of the wave-vector in Cartesian coordinates
\be
k_x(t) = \tilde{k}_{x} - A t \tilde{k}_z, \,\, k_y =
\tilde{k}_y, \,\, k_z= \tilde{k }_{z}.
\label{k(t)}
\ee
This time dependence of the wave vector $\mathbf{k}(t)$ has purely
kinematic origin and it is not related to the dispersion of the
waves. Such a phenomenon is well-known in the acoustics of moving
media, but it can affect even non-propagating spatial structures
with ``tagged'' atoms. Even in this static case with exactly zero
frequency the general formula for the wave-vector evolution
\Eqref{k(t)} is applicable.

In the next section we will apply these kinematic relations to the
Alfv\'{e}n waves. We consider the Alfv\'{e}n velocity $V_A,$ defined
by an external magnetic field $B_0,$ the density of the fluid $\rho$
and magnetic pressure $p_{_B}$
\be
\label{pressuremagnitno}
p_{_B}=\frac{B_0^2}{2\mu_0}=\frac{1}{2}\rho V_A^2,
\ee
and the shear parameter $A$ with dimension of a frequency. In
Gaussian system the magnetic permeability of the vacuum is
$\mu_0 = 4\pi;$ in SI for simplicity
$\mu_0 = 4\pi\times 10^{-7}.$
For a kinematic description it is convenient to introduce
dimensionless wave vectors
$\mathbf{K} = (V_A/A)\, \mathbf{k}, \quad \tilde{\mathbf{K}} =
(V_A/A)\, \tilde{\mathbf{k}},$
as well as dimensionless time $\tau = A t$; $L_A\equiv V_A/A$ is the
unit for length. The connection between the Eulerian components of
the wave-vector and the components in the tilde Lagrangian system
\Eqref{k(t)} reads as
\begin{align}
\label{Kx(tau)}
 K_{x}(\tau) = \tilde{K}_{x} (0) - \tau \tilde{K}_{z}, \nn\\
 K_{y}(\tau) = \tilde{K}_{y} (0) , \quad K_{z}(\tau) =
\tilde{K}_{z} (0) .
\end{align}

With an appropriate choice of the initial time we can further set
$\tilde{K}_{x} (0)=0.$ Here we have used that the wave-vector in the
Lagrangian (tilde in our notations) coordinate system is constant,
$\tilde{\mathbf{K}}(\tau)=\mathrm{const}.$ Lagrangian coordinates
are also known as flow-following coordinates, which in this
particular flow preserve the wave vector (similarly, a plane of
``tagged'' atoms is moved but not deformed). For matrix presentation
of the considered relations see Appendix~\ref{app:matrixpresent}.

\section{Linearized stochastic MHD}\label{sec:stochMHD}

Consider the laminar component of the velocity field
$\mathbf{V}(t,\mathbf{r})$ of an incompressible flow of an accreting
plasma with a constant density $\rho=\mathrm{const}.$ In the
presence of an external magnetic field $\mathbf{B}(t,\mathbf{r})$
the velocity $\mathbf{V}(t,\mathbf{r})$ evolves according to the
Navier-Stokes equation with a Lorentz force
\be \label{Navier-Stokes}
\rho (\partial_t  + \mathbf{V} \cdot \nabla)\mathbf{V}
= - \nabla p + \eta \Delta \mathbf{V}
  + \mathbf{j} \times \mathbf{B}
  + \mathbf{F}, \ee
where $p$ is the pressure and $\eta$ is the viscosity
\cite{L.VIII}. The term $\mathbf{F}(t,\mathbf{r})$
phenomenologically describes the Reynolds and Maxwell stresses
associated with the turbulence. We will return to this term later,
when we analyze the evolution of perturbations in the shear flow.

The current $\mathbf{j}$ is given by Ohm's law
\be
\label{Ohm2}
\mathbf{j} = \sigma(\mathbf{E} + \mathbf{V}\! \times \mathbf{B}) \, ,
\ee
where $\sigma\equiv 1/\varrho$ is the electrical conductivity and
$(\mathbf{E} + \mathbf{V}\! \times \mathbf{B})$ expresses the
effective electric field acting on the fluid; the velocity $V\ll
c$ is nonrelativistic.

For low frequencies $j \gg \varepsilon_0|\partial_t \mathbf{E}|$,
so it is possible to use the magnetostatic approximation
\be
\label{Maxeqn} \mathrm{rot}\,
\mathbf{B} = \mu_0 \mathbf{j}\, .
\ee
We substitute here the current from \Eqref{Ohm2} and obtain
 \be
 \mathbf{E} = - \mathbf{V} \times \mathbf{B}
 + \nu_m \, \mathrm{rot}\, \mathbf{B},
\ee
where $\nu_m\equiv \varepsilon_0 c^2\varrho$ is the magnetic viscosity.
In Gaussian system $\varepsilon_0 = 1/4\pi$ while in SI
$\varepsilon_0 = 1/(4\pi\times 10^{-7}c^2).$

Here we will remind some basic properties of the plasma. We suppose
that the frequency of ion-ion collisions $\nu_{ii}$ is much bigger
than the ion cyclotron frequency $\omega_{B_i}$
\be \nu_{ii} \gg \omega_{B_i},\quad \omega_{B_i}=eB/M, \ee
where $M$ is the mass of the ion, see Ref.~[\onlinecite{L.X}],
secs.: 41-43, 58. For weak fields we can neglect the influence of
the magnetic field over the viscosity and resistivity; see
Ref.~[\onlinecite{L.X}],~eqs.~(43.8-43.10)
\begin{eqnarray}
\label{ocenkazakinvisc}
\nu_k & \approx & \frac{0.4\,T_p^{5/2}}{e^4N_pM^{1/2}\mathcal{L}_p},
\quad \frac{\sigma}{4\pi\varepsilon_0}\approx\frac{0.6T_e^{3/2}}{e^2m^{1/2}\mathcal{L}_e}   \\
\label{ocenkaZaMagnVisc} \nu_m & \approx &
\frac{e^2c^2m^{1/2}\mathcal{L}_e}{0.6*4\pi\,T_e^{3/2}},\quad
e^2 \equiv q_e^2/4\pi\varepsilon_0,
\end{eqnarray}
where $m$ is the electron mass, $q_e$ is the electron charge, $N$
is the number of electrons per unit volume, $\mathcal{L}$ is the
Coulomb logarithm and $r_D$ is the Debye length
\begin{eqnarray}
\nn \mathcal{L} &= &\ln\left(\frac{r_DT}{e^2}\right)=\ln\left(4\pi r_D^3N\right),\\
r_D &= & \sqrt{\frac{T}{4\pi e^2N}}.
\end{eqnarray}
For high enough temperatures $T\gg T_{km},$ where
\be \label{tempNaIzravnqvane} T^4_{km}=4\pi
mc^2e^6N\mathcal{L}^2\sqrt{\frac{M}{m}}, \ee
the kinematic viscosity dominates $\nu_k\gg\nu_m .$

The evolution of the magnetic field is governed by the other Maxwell
equation $\mathrm{rot}\, \mathbf{E} = - \partial_t \mathbf{B}$, so
that the induction equation reads
 \be
 \partial_t \mathbf{B} = \mathrm{rot}\,(\mathbf{V} \times \mathbf{B}) -
  \nu_m\mathrm{rot}\,\mathrm{rot}\,\mathbf{B}.
 \ee
For an incompressible flow $\mathrm{div}\,\mathbf{V} \!\!= \!0,$
taking into account that $\mathrm{div}\,\mathbf{B}\!\! = \!0,$ the
equation above takes the form
\be
\label{Bdynamics}
(\partial_t + \mathbf{V}\! \cdot \! \nabla) \mathbf{B} =
(\mathbf{B}\! \cdot \! \nabla) \mathbf{V} + \nu_m\Delta \mathbf{B} \, ,
\ee
where we have used the general relation
$$
\mathrm{rot} (\mathbf{V} \times \mathbf{B})=
 \mathbf{V} \mathrm{div} \mathbf{B}
+ (\mathbf{B} \cdot \nabla) \mathbf{V}
-\mathbf{B} \, \mathrm{div} \mathbf{V}
- (\mathbf{V} \cdot \nabla) \mathbf{B}.
$$
We will use the dynamics equations Eqs.~(\ref{Navier-Stokes},
\ref{Bdynamics}) to analyze the propagation of MHD waves in a shear
flow. Let the velocity $\mathbf{V}$ be a sum of equilibrium shear
velocity $\mathbf{U}^{(0)}(\mathbf{r})$ and a small perturbation
$\mathbf{u}(t,\mathbf{r}),$ for which we are going to derive
linearized wave equations
\be
\mathbf{V} = \mathbf{U}^{(0)} + \mathbf{u}.
\ee
The same representation we suppose for the magnetic
field and the pressure
\be
\quad \mathbf{B} = \mathbf{B}_0 + \mathbf{B}^{'},\quad
p = p_0 + p^{'}.
\ee
As mentioned in section \ref{sec:Kinematics}, the $z$-axis of the
coordinate system we choose along the shear velocity and $x$-axis
along the velocity gradient
 \be
 \mathbf{U}^{(0)}= (0, 0, Ax)
 = A x \, \mathbf{e}_z,\quad \mathbf{e}_z=(0,0,1).
 \ee
For this choice of the coordinates the vorticity
\be
\mathrm{rot}\, \mathbf{U}^{(0)}=(0,-A,0)=-A\mathbf{e}_y,
\quad \mathbf{e}_y= (0,1,0),
\ee
is along $y$-axis.

The differential rotation in the accretion disks stretches the
frozen-in magnetic field lines and eliminates the cross-flow
magnetic field. In the present work we will consider the magnetic
field in the plasma to be parallel to the shear flow in z-direction
\be
\mathbf{B}_0 = (0, 0, B_0)= B_0\mathbf{e}_z.
\ee

We will consider a general perturbation case
\be
\mathbf{u}(t,\mathbf{r}) = (u_x, u_y, u_z),\quad
\mathbf{B}^{'}(t,\mathbf{r}) = (B^{'}_x,B^{'}_y,B^{'}_z).
\ee
Neglecting the small quadratic terms $(\mathbf{u}\cdot \nabla)
\mathbf{u}$ and $(\mathbf{u}\cdot \nabla) \mathbf{B}^{'}$ in the
substantial derivatives in Eqs.~(\ref{Navier-Stokes},
\ref{Bdynamics}) the linearized evolution equations read (see
Appendix~\ref{app:dyneqs})
\begin{eqnarray}\label{Euler}
(\partial_t + Ax\partial_z)
\mathbf{u}
= -\frac{\nabla p^{'}}{\rho} -
Au_x\mathbf{e}_z
+\frac{\mathbf{F}}{\rho}
\nn \\
 + \frac{B_0}{\mu_0 \rho}\!\! \left(
\begin{matrix}\partial_z B^{'}_x -\partial_x B^{'}_z\\
\partial_z B^{'}_y  - \partial_y B^{'}_z\\
0
\end{matrix}
\right) + \nu_k \Delta\mathbf{u},\\
(\partial_t + Ax\partial_z)
\mathbf{B}^{'}
= B_0 \partial_z
\mathbf{u}
\nn \\
+ AB^{'}_x \mathbf{e}_z + \nu_m \Delta \mathbf{B}^{'}.
\end{eqnarray}
For the density of the random force that models the turbulence we
assume a white noise correlator
\be \langle
\mathbf{F}(t_1,\mathbf{r}_1)\mathbf{F}(t_2,\mathbf{r}_2)\rangle
=\tilde\Gamma{\rho}^2\delta(t_1-t_2)\delta(\mathbf{r}_1-\mathbf{r}_2)\openone,
\ee
parameterized by the Burgers parameter $\tilde\Gamma.$
In such a way we derive in Eulerian coordinates $\mathbf{r} =
(x,y,z)$ a system of partial differential equations. The transition
to Lagrangian $\tilde{\mathbf{r}} = (\tilde{x},\tilde{y},\tilde{z})$
variables \Eqref{newvar}, however, reduces it to a system of
ordinary differential equations. To \Eqref{newvar} we add also
$\tilde{t}\equiv t$ and perform the change of variables
\be
\mathbf{u}(t,\mathbf{r})=\mathbf{u}(\tilde{t},\tilde{\mathbf{r}}),\quad
\mathbf{B}^{'}(t,\mathbf{r})=\mathbf{B}^{'}(\tilde{t},\tilde{\mathbf{r}}),\quad
\mathbf{F}(t,\mathbf{r})=\mathbf{F}(\tilde{t},\tilde{\mathbf{r}}),
\ee
supposing that $(\tilde{t},\tilde{\mathbf{r}})$ are 4 independent
variables as $(t,\mathbf{r})$ are. From \Eqref{newvar} one can
easily derive the rules for the change of variables in the
derivatives
\be
\partial_t = \partial_{\tilde{t}} - A\tilde{x}\partial_{\tilde{z}},\;\,
\partial_x = \partial_{\tilde{x}} - A\tilde{t}\partial_{\tilde{z}},\;
\partial_y = \partial_{\tilde{y}},\;
\partial_z = \partial_{\tilde{z}}.
\ee
The main advantage of this transition to Lagrangian variables is
that the substantial time derivative
$\partial_t + Ax\partial_z=\partial_{\tilde{t}}$
in the MHD set of equations \Eqref{Euler} no longer depends on the
spatial coordinates.
\section{Separation of variables}
\label{sec:SepOfVar}
For a system with constant (space independent) coefficients the
solutions of the equations are plane waves. That is why using the
phase invariance \Eqref{phase-invariance} one can seek a solution in
the form
\begin{eqnarray}
\label{plane-wave}
\mathbf{u} =
\mathbf{u}_{\tilde{\mathbf{k}}}(\tilde{t})\exp(i\tilde{\mathbf{k}}
\cdot \tilde{\mathbf{r}})
= \mathbf{u}_{\tilde{\mathbf{k}}}(t)\exp(i\mathbf{k}(t) \cdot \mathbf{r}),\\
\mathbf{B}^{'}
= \mathbf{B}^{'}_{\tilde{\mathbf{k}}}(\tilde{t})\exp(i\tilde{\mathbf{k}} \cdot \tilde{\mathbf{r}})
= \mathbf{B}^{'}_{\tilde{\mathbf{k}}}(t)\exp(i\mathbf{k}(t) \cdot \mathbf{r}),
\end{eqnarray}
where
$\mathbf{u}_{\tilde{\mathbf{k}}}(\tilde{t})=\mathbf{u}_{\tilde{\mathbf{k}}}(t)$ and
$\mathbf{B}^{'}_{\tilde{\mathbf{k}}}(\tilde{t})=\mathbf{B}^{'}_{\tilde{\mathbf{k}}}(t)$
are the time dependent amplitudes of the plane waves.
For now we suppose that the variables are complex-valued.

For clarity we will perform the change of variables gradually and
will analyze the result for each term individually. First we will
change the variables only in the substantial time derivatives so
that the system \Eqref{Euler} takes the form
\begin{eqnarray}
\label{Euler-Langange}
\partial_{\tilde{t}}
\mathbf{u}
+
A u_x \mathbf{e}_z
-\frac{\mathbf{F}}{\rho}
=\nn\\
 - \!\frac{\nabla p^{'}}{\rho} + \frac{B_0}{\mu_0 \rho}\!\!
\left(
\begin{matrix}
\partial_{z} B^{'}_{x} - \partial_{x} B^{'}_{z} \\
\partial_{z} B^{'}_y  - \partial_{y} B^{'}_z \\
0
\end{matrix}
\right)\! + \nu_k \Delta \mathbf{u},\\
\label{Bstart}
\partial_{\tilde{t}}
\mathbf{B}^{'}
= \! B_0 \partial_{z}
\mathbf{u} + AB^{'}_x\mathbf{e}_z
+ \nu_m \Delta \mathbf{B}',\\
\partial_{x}B^{'}_x +
\partial_{y}B^{'}_y +
\partial_{z}B^{'}_z = 0,
\label{div_B}
\\
\partial_{x}u_x +
\partial_{y} u_y+
\partial_{z} u_z = 0.
\label{div_u}
\end{eqnarray}
To emphasize the physical meaning we have used Lagrangian variables,
however one can start with the simple relation which both automatizes and ensures
the separation of variables
\begin{eqnarray}
\label{razdNaProm}
(\partial_t&+&Ax\partial_z)
\left(\mathbf{u}(t)
 \exp \left\{ i \left[
(\tilde{k}_x-At\tilde{k}_z)x +\tilde{k}_y y +\tilde{k}_z z
                \right]
      \right\}
\right)\nn\\
&=&\exp \left\{ i \left[
(\tilde{k}_x-At\tilde{k}_z)x +\tilde{k}_y y +\tilde{k}_z z
                \right]
      \right\}
\mathrm{d}_t\mathbf{u}(t).
\end{eqnarray}
For the plane-wave in \Eqref{plane-wave} the
substantial derivatives
\be
D_t\equiv \partial_t+\mathbf{V}\cdot\nabla
\ee
after linearization take the form
\begin{eqnarray}
D_t\left[\mathbf{u}_{\tilde{\mathbf{k}}}(t)
 \exp(i\mathbf{k}(t) \cdot \mathbf{r})\right]\nn\\
\approx\exp(i\mathbf{k}(t) \cdot \mathbf{r})
 \left[\mathrm{d}_{\tilde{t}}\mathbf{u}_{\tilde{\mathbf{k}}}(t)
 +u_{z,\tilde{\mathbf{k}}}(t)\mathbf{e}_z
 \right], \label{sd_lin_u}\\
 D_t\left[\mathbf{B}^{'}_{\tilde{\mathbf{k}}}(t)
 \exp(i\mathbf{k}(t) \cdot \mathbf{r})\right]
\approx\exp(i\mathbf{k}(t) \cdot \mathbf{r})
 \mathrm{d}_{\tilde{t}}\mathbf{B}^{'}_{\tilde{\mathbf{k}}}(t).\label{sd_lin_B}
\end{eqnarray}
As mentioned earlier, in  Eqs.~(\ref{sd_lin_u}, \ref{sd_lin_B}) we
have neglected the quadratic terms, which describe a wave-wave
interaction.

Now we can substitute here the right-hand side of the supposed
plane form of the waves from \Eqref{plane-wave}, where they are
presented in Eulerian coordinates. The latter are more convenient
to calculate the gradients $\nabla=i\mathbf{k}(\tilde{t})$ and
Laplacians $\Delta=- k^2(\tilde{t})$ but we have to remember that
according to \Eqref{k(t)} the wave-vectors
$\mathbf{k}(\tilde{t})=(A/V_A)\mathbf{K}(\tau\!=\!A\tilde{t})$ are
time dependent
\be
\label{kEuler}
\mathbf{k}(\tilde{t})= (
  \tilde{k}_{x} - A \tilde{t} \tilde{k}_z, \,\,
  \tilde{k}_y, \,\,
  \tilde{k }_{z}       ), \quad
k^2(\tilde{t})=k^2_x+k_y^2+k_z^2.
\ee
Additionally, to obtain a system of ordinary differential equations
for the amplitudes of the MHD waves, one can eliminate the pressure
as it is done in Appendix~\ref{app:MHDsyst}. We suppose plane waves
for the velocity and the magnetic field
\begin{eqnarray}
\label{u_plane_wave}
\mathbf{u}(t,\mathbf{r})=\mathbf{u}_{\tilde{\mathbf{k}}}(t)
\exp(i\mathbf{k}(t)\cdot\mathbf{r}),\\
\label{B_plane_wave}
\mathbf{B}'(t,\mathbf{r})=\mathbf{B}^{'}_{\tilde{\mathbf{k}}}(t)
\exp(i\mathbf{k}(t)\cdot\mathbf{r}),
\end{eqnarray}
where we have introduced several notations that will be used later
\begin{align}
\label{bezrazmSkorost}
\mathbf{u}_{\tilde{\mathbf{k}}}(t)
=-i\tilde{\mathbf{u}}_{\tilde{\mathbf{k}}}(t)
=-iV_A\vec{\upsilon}_{\tilde{\mathbf{k}}}(\tau),\\
\label{bezrazmMagnP}
\mathbf{B}^{'}_{\tilde{\mathbf{k}}}(t)
=B_0\mathbf{b}_{\tilde{\mathbf{k}}}(\tau).
\end{align}
For the sake of brevity the Lagrange wave-vector indices
$\tilde{\mathbf{k}}$ will be further suppressed; for example, taking
the real part from Eqs.~(\ref{u_plane_wave}, \ref{B_plane_wave}) we
have
\begin{eqnarray}
\label{u_plane}
\mathbf{u}(t,\mathbf{r})=V_A\vec{\upsilon}(\tau)
\sin(\mathbf{k}(t)\cdot\mathbf{r}),\\
\label{B_plane}
\mathbf{B}'(t,\mathbf{r})=B_0\mathbf{b}(\tau)
\cos(\mathbf{k}(t)\cdot\mathbf{r}).
\end{eqnarray}

For the real dimensionless amplitudes $\vec{\upsilon}$ and
$\mathbf{b}$ after some algebra given in the
Appendix~\ref{app:MHDsyst} we derive the system of equations:
\begin{eqnarray}
\mathrm{d}_{\tau}\upsilon_x &=&\frac{2K_zK_x}{K^2}\upsilon_x
-K_zb_x-\nu_k'K^2\upsilon_x
+gf_x,\\
\mathrm{d}_{\tau}\upsilon_y &=&\frac{2K_zK_y}{K^2}\upsilon_x
-K_zb_y-\nu_k'K^2\upsilon_y
+hf_y,\\
\mathrm{d}_{\tau} b_x &=& K_z \upsilon_x - \nu_m^{\prime} K^2(\tau) b_x,\\
\mathrm{d}_{\tau} b_y &=& K_z \upsilon_y - \nu_m^{\prime} K^2(\tau) b_y,\\
\label{pulnanesviv}
\upsilon_z &=& - \frac{K_x}{K_z}\upsilon_x - \frac{K_y}{K_z}\upsilon_y ,\\
b_z &=& - \frac{K_x}{K_z}b_x - \frac{K_y}{K_z}b_y,
\end{eqnarray}
where
\begin{align}
g(\tau)=\frac{\sqrt{K_y^2+K_z^2}}{K(\tau)},\quad
h(\tau)=\frac{\sqrt{K_x^2(\tau)+K_z^2}}{K(\tau)},\\
\label{ratio}
K_x=-\tau K_z,\quad K_y=\mathrm{const},\quad K_z=\mathrm{const},\\
K(\tau)=\sqrt{K_x^2+K_y^2+K_z^2}=\sqrt{K_y^2+(1+\tau^2)K_z^2},\\
\nu_k^{\prime}\equiv \frac{A}{V^2_A} \nu_k,\quad
\label{bezrazmMagnVisc}
\nu_m^{\prime}\equiv \frac{A}{V^2_A} \nu_m,\quad
\Gamma=\frac{\tilde{\Gamma}}{V_A^2A\mathcal{V}},\\
\label{pulenMagnVisc}
\nu\equiv \nu_k+\nu_m,\quad \nu'\equiv \nu_k'+\nu_m',\\
\langle f_x(\tau_1) f_x(\tau_2)\rangle
=\langle f_y(\tau_1) f_y(\tau_2) \rangle
=\Gamma\delta(\tau_1-\tau_2),
\end{align}
and $\cal{V}$ is the volume of the system.

In the simplest case of zero shear $A=0$ and dissipation $\nu_k=
\nu_m = 0$ we have a system with constant coefficients which gives
in dimensionless form the dispersion of Alfv\'en
waves~[\onlinecite{L.VIII}], sec.~69, problem
\begin{eqnarray}
\omega_A (\mathbf{k})=\frac{V_A}{B_0}
\left| \mathbf{k}\cdot\mathbf{B}_0 \right|
-\frac{i}{2}\nu\mathbf{k}^2,\quad
V_A \equiv \frac{B_0}{\sqrt{\mu_0\rho}},\\
\mathbf{V}_{\mathrm{gr}}(\mathbf{k})
=\frac{\partial \omega_A (\mathbf{k})}{\partial\mathbf{k}}
=V_A \, \mathrm{sign}(\mathbf{k}\cdot\mathbf{B}_0 )\, \frac{\mathbf{B}_0}{\mathrm{B}_0}.
\end{eqnarray}
The wave-wave interaction is small and negligible only if the
dimensionless wave components of the velocity and magnetic fields
are sufficiently small $\upsilon^2,\; b^2 \ll1.$
\section{Two dimensional waves}
\label{sec:2Dwaves}
To investigate the influence of the shear flow on the evolution of
the MHD waves, here we will concentrate on the two-dimensional case
$k_y=0$ which gives a complete separation of the variables. For this
special case $K_y=0$ the dynamic equations take the form
\begin{eqnarray}
\label{upsi=}
\mathrm{d}_{\tau}\upsilon_x
 &=& \alpha(\tau)\upsilon_x
 - K_z b_x\nn\\
 && - \nu_k' K^2(\tau) \upsilon_x
 + g(\tau)f_x(\tau),\\
\label{upsiy=} \mathrm{d}_{\tau} \upsilon_y &=& - K_z b_y -
\nu_k^{\prime} K^2(\tau)\upsilon_y
+f_y(\tau),\\
\label{bex=}
\mathrm{d}_{\tau} b_x &=& K_z \upsilon_x - \nu_m^{\prime} K^2(\tau) b_x,\\
\label{bey=}
\mathrm{d}_{\tau} b_y &=& K_z \upsilon_y - \nu_m^{\prime} K^2(\tau) b_y,\\
\label{nesvivaemost}
\upsilon_z &=& - \frac{K_x}{K_z}\upsilon_x ,\\
b_z &=& - \frac{K_x}{K_z}b_x ,
\end{eqnarray}
where
\begin{eqnarray}
\label{alpha}
\alpha(\tau)
&\equiv& 2 \frac{K_x(\tau)K_z}{K^{2}_x(\tau)+ K^{2}_z}
 =-\frac{2\tau}{1+\tau^2},\\
K^2(\tau)&=&(1+\tau^2)K_z^2,\quad
g(\tau)= \frac{1}{\sqrt{1+\tau^2}}.
\end{eqnarray}

We will start our analysis with the case of short wavelengths in a
dissipationless and a fluctuation-free regime.  For a highly
conducting plasma $\nu_m \approx 0$ with a negligible kinematic
viscosity $\nu_k \approx 0$ (conditionally, we may say superfluid
and superconducting plasma) the system above yields
\begin{eqnarray}
\label{deu_x}
 \mathrm{d}_{\tau} \upsilon_x
  &= &\alpha(\tau)\, \upsilon_x - K_z b_x,\\
\label{deb_z}
 \mathrm{d}_{\tau} b_x &=& K_z \upsilon_x.
\end{eqnarray}
We differentiate \Eqref{deu_x} with respect to time, neglect (for
$|K_z|\gg1$) the $\mathrm{d}_{\tau}\alpha$ term, and substitute
$\mathrm{d}_\tau b_x$ from \Eqref{deb_z} which implies
\be
\label{starteq}
\ddot{\upsilon} -
\alpha(\tau)\dot{\upsilon} + K_z^2 \upsilon \approx 0,
\ee
where $\upsilon \equiv \upsilon_x$ and the dot stands for the
dimensionless time $\tau$ derivative.
 In the original variables this equation reads
\be
\label{oscillator}
\mathrm{d}_t^2 u_{x,\mathbf{\tilde{k}}} +
\overline\gamma_s(t) \mathrm{d}_t u_{x,\mathbf{\tilde{k}}} + \omega_A^2
u_{x,\mathbf{\tilde{k}}}\approx 0,
\ee
where the frequency of the Alfv\'en waves
\be
\omega_A(\mathbf{k})=V_A|k_z|=V_Ak|\cos\varphi|,
\ee
depends on the angle
\be
\label{angle_kB}
\varphi(t)=\arccos\frac{\mathbf{k}\cdot\mathbf{B}_0}{kB_0}
=\arctan\frac{k_x(t)}{k_z}=-\arctan \tau
\ee
between the external magnetic field and the wave-vector. In such a
way we find the geometrical meaning of the dimensionless time.

The effective friction coefficient $\overline\gamma_s(t)$ in the
oscillator equation \Eqref{oscillator} is presented by the
dimensionless function $\alpha(\tau)$
\begin{eqnarray}
\overline\gamma_s(t)&=&A\gamma_s(\tau), \\
\gamma_s(\tau)&\equiv&-\alpha(\tau)=\frac{2\tau}{1+\tau^2}
=\sin2\varphi.
\end{eqnarray}
Without any restrictions we have considered $\tilde{k}_x=0$ and $K_x=0$
because it is related only to the choice of the initial time \Eqref{kEuler}.

Our next step is to determine the conditions that ensure small
dissipation rate. Ohmic resistivity and viscosity produce an
additional friction coefficient in the effective oscillator
equation \Eqref{oscillator}; see Ref.~[\onlinecite{L.VIII}],
sec.~69, problem
\begin{eqnarray}
\label{DissipativeDecrement}
\overline\gamma_{\nu}= (\nu_k+\nu_m) \mathbf{k}^2(t)= A\gamma_{\nu},\\
\gamma_{\nu}(\tau)=(1+\tau^2)\,a
=\frac{a}{\cos^2\varphi},\\
\label{bezrazmernaskorostnazatihvane}
a\equiv \nu'K_z^2=(\nu_k+\nu_m)k_z^2/A.
\end{eqnarray}
The upper formulae are meticulously obtained in
Appendix~\ref{app:koefnazatihv}.
%
%
Taking into account the small dissipation in the oscillator equation
(\ref{oscillator}) we have to substitute the shear attenuation with
the total attenuation
\be
\label{decrement}
\overline\gamma
=\overline\gamma_s(\varphi)+\overline\gamma_{\nu}
=A\gamma=A(\gamma_s+\gamma_{\nu}).
\ee
The dimensionless attenuations $\gamma_s$ and $\gamma_{\nu}$ are
more convenient for further analysis of the kinetics of Alfv\'{e}n
waves.

From \Eqref{oscillator} we derive an equation for the attenuation of
the averaged effective energy for a fictitious particle with a
coordinate $u_{x,\mathbf{\tilde{k}}}$
\be \overline{E}_{\mathrm{eff}}(t)
=\frac{1}{2}\left(\dot{u}_{x,\mathbf{\tilde{k}}}^2 +\omega_A^2
u_{x,\mathbf{\tilde{k}}}^2\right) \ee of the oscillator
[\onlinecite{L.I}],~Eq.~(25.5), ibid.~[\onlinecite{L.I}], sec.~51,
problem~2
\be
\label{EnergyKinetics}
\mathrm{d}_t \overline{E}_{\mathrm{eff}}\approx
-\overline\gamma(\varphi)\overline{E}_{\mathrm{eff}}.
\ee
Here the time dependence is implicitly included by the dependence
of the attenuation on the angle $\varphi,$ \Eqref{angle_kB}.

To derive the approximate oscillator equation \Eqref{starteq} we
assumed that the alteration of the wave vector components $K(\tau)$
and respectively the variation of $\alpha(\tau)$ (or $\gamma_s(t)$
in \Eqref{oscillator}) is negligible compared to the change of the
velocity $\upsilon$ and its derivatives and therefore all terms
containing $\dot{K_x}(\tau)$ and $\dot{\alpha}(\tau)$ have been
omitted. Such approximation corresponds to the WKB approximation in
quantum mechanics where we have fast-oscillating amplitude and
slowly changing coefficients. The approximate equation
\Eqref{oscillator} will be the starting point for our further
analysis.

We will illustrate the WKB method for calculation of the
attenuation on the simplest possible example of the system
\Eqref{deu_x} and \Eqref{deb_z} or equation \Eqref{starteq}.
Substitution in this equation of amplitudes $\propto \exp(\lambda
\tau)$ gives the characteristic equation
\begin{eqnarray}
D(\lambda)=P(\lambda)+V(\lambda)=0,\quad\\
P(\lambda)=\lambda^2+K_z^2,\quad
\mathrm{d}_\lambda P(\lambda)=2\lambda,\quad
V(\lambda)=\gamma_s \lambda.
\end{eqnarray}
Supposing $\gamma_s$ to be a small perturbative correction we have
in the first Newtonian iteration
\begin{eqnarray}
P(\lambda_0)=0,\quad
\lambda_0=-i K_z,\quad\\
\label{firstNewtonIteration}
\lambda\approx
\lambda_0-\frac{V(\lambda_0)}
           {\mathrm{d}_\lambda P(\lambda_0)}
= -i K_z - \frac{1}{2}\gamma_s.
\end{eqnarray}
In such a way for the energy of the wave we obtain
\be
\overline{E}(\tau)\propto |\exp(\lambda\tau)|^2
=\exp(-\gamma_s\tau)
=\exp(-\overline\gamma_s t).
\ee
This procedure applied to the more general system of equations
Eqs.~(\ref{upsi=}, \ref{bex=}) leads to the secular equation
\begin{equation}
\left|
\begin{matrix}
\gamma_s + \nu_k^\prime K^2(\tau) + \lambda& -K_z \\
K_z &\nu_m^\prime K^2(\tau) + \lambda
\end{matrix}
\right|
= 0.
\end{equation}
Thus the general expression for the eigenvalues is
\begin{equation}
\lambda^2 + \gamma\lambda + K_z^2
+ \left[\gamma_sK^2(\tau) + \nu_k^{\prime}K^4(\tau)\right]\nu_m^\prime =0.
\end{equation}
For high enough temperatures $T \gg T_{km}$ the last term is
negligible (see Eqs.~(\ref{ocenkazakinvisc}, \ref{ocenkaZaMagnVisc}
and \ref{tempNaIzravnqvane})). Then we may ignore the slight
variation of the real part of the Alfv\`{e}n frequency and account
only for the imaginary correction
\begin{equation}
\lambda \approx - iK_z - \frac{1}{2}\gamma,
\end{equation}
which after a time differentiation gives the kinetic equation for
the averaged energy of the MHD wave \Eqref{EnergyKinetics}.

The damping of the waves can be calculated directly as a ratio of
the dissipated power divided by the energy. Initially, we have to
take the real part of the wave variables and then we have to average
over the period of oscillations
\be
\label{diss_rate}
\overline\gamma_\nu=
\frac{
\langle
\mathbf{j}(t,\mathbf{r})\cdot \mathbf{E}(t,\mathbf{r})
\rangle
+\frac{\eta}{2}
\sum_{i,k=1}^{3}\langle\left(\partial_i u_k+\partial_k u_i\right)^2\rangle}
{\frac{\rho}{2}\langle \mathbf{u}^2(t,\mathrm{r})\rangle
+\frac{1}{2\mu_0}
\langle \mathbf{B}^{\prime \, 2}(t,\mathrm{r})\rangle},
\ee
where the numerator represents the volume density of the total
dissipated power $Q_\mathrm{tot}$ and
\be
\partial_k\equiv\frac{\partial}{\partial x_k}, \qquad (x_1,\,x_2,\,x_3)=(x,\,y,\,z).
\ee
\Eqref{diss_rate} is an alternative way to derive
\Eqref{DissipativeDecrement} for the case of small shear as it is
done in Appendix~\ref{app:koefnazatihv}.

\section{Kinetics of Alfv\'{e}n amplitudes in WKB approximation}
\label{sec:WKB}
In the spirit of quantum mechanics the number of
particles is proportional to the square of the amplitude. In this
sense using the dimensionless amplitude of the velocity $\upsilon$
one can introduce the mean number of ``alfvons''
\be
\label{n=sredno(v^2)}
n(\tau)=\langle (\mathrm{Re}\, \upsilon)^2 \rangle,
\ee
where the time $\tau$ average is taken over the dimensionless period
of the Alfv\'{e}n waves $2\pi A/\omega_A=2\pi/|K_z|.$ According to
the definition Eqs.~(\ref{u_plane}, \ref{B_plane}) the variables
$\vec{\upsilon}$ and $\mathbf{b}$ are real, so the sign for a real
part can be omitted.

The volume density of the wave energy is proportional to the square
of the amplitude and the number of ``alfvons''
\be
\overline E
=2\,\frac{\rho}{2}
\langle ( \mathrm{Re} \, u_x)^2 + (\mathrm{Re} \, u_z)^2\rangle
= \rho V_A^2 (1 + \tau^2)n.
\ee
Here we have taken into account that the averaged kinetic energy of
the fluid is equal to the averaged energy of the magnetic field
which plays the role of the ``potential'' elastic energy of these
transversal waves. We have also used the explicit form of the
wave-vectors Eqs.~(\ref{ratio}, \ref{nesvivaemost}). The influence
of the $y$ mode will be assessed later. In such an interpretation
the equation for the time derivative of the wave energy
\Eqref{EnergyKinetics} can be considered as a Boltzmann kinetic
equation for the number of ``alfvons''
\begin{eqnarray}
\label{Boltzmanneq.}
d_{\tau} n(\tau)&=& -\gamma(\tau)n(\tau)+w(\tau),\\
\gamma&=&\gamma_s+\gamma_\nu,\\
\gamma_s&=&\frac{2\tau}{1+\tau^2},\\
\gamma_\nu&=&(1+\tau^2)a.
\end{eqnarray}
In order not to interrupt the explanation the derivation of the
turbulence-induced source term $w$ will be given in a separate
subsection~\ref{GeneralTheorem}, see~\Eqref{www}.

One can easily check that the solution of the kinetic
equation \Eqref{Boltzmanneq.} reads
\be
\label{solution}
n(\tau)=\int_{-\infty}^{\tau}w(\tau^{\prime})\exp
\left[
-\int_{\tau^{\prime}}^{\tau}\gamma(\tau^{\prime\prime})
 \mathrm{d}\tau^{\prime\prime}
\right]
\mathrm{d}\tau^{\prime}.
\ee

Let us first analyze in \Eqref{Boltzmanneq.} the dissipationless
regime of $\gamma_\nu=0$ and zero turbulence power $w=0.$
Using the integral
\be
\exp\left(-\int_{0}^{\tau}\gamma_s(\tau^{\prime\prime})
 \mathrm{d}\tau^{\prime\prime}\right)
=\cos^2 \varphi
=\frac{1}{1+\tau^{2}},
\ee
for the solution of the homogeneous linear equation we obtain
\be
\label{pribl.br.chast.}
n(\tau)\approx n_0\cos^2 \varphi=\frac{n_0}{1+\tau^{2}},
\ee
where the integration constant $n_0$ determines the spectral density
of the waves with wave vector $\mathbf{k}$ parallel to the constant
external magnetic field $\mathbf{B}_0$.

Further, in the general solution \Eqref{solution} we have to
substitute the integral
\be
\exp\left(-\int_{\tau^{\prime}}^{\tau}\gamma_s(\tau^{\prime\prime})
 \mathrm{d}\tau^{\prime\prime}\right)
=\frac{\cos^2 \varphi }{\cos^2 \varphi^{\prime}}
=\frac{1+\tau^{\prime 2}}{1+\tau^{2}}.
\ee
The contribution of the dissipation is given by
\be
\int_{0}^{\tau}\gamma_{\nu}(\tau^{\prime\prime})
 \mathrm{d}\tau^{\prime\prime}
=\left(\tau + \frac{1}{3}\tau^3 \right)a,
\ee
or for the interval pointed above \Eqref{solution}
\be
-\int_{\tau^{'}}^{\tau}\gamma_{\nu}(\tau^{\prime\prime})
 \mathrm{d}\tau^{\prime\prime}
 = a \left[-(1+\tau^2)u+\tau u^2-\frac{1}{3}u^3\right],
\ee
where for simplicity we have introduced a shifted time variable
\be
u\equiv \tau-\tau^{\prime}>0.
\ee
Finally, the solution of the Boltzmann equation \Eqref{solution}
takes the form
\begin{eqnarray}
\label{BashSolution}
n(\tau)=
\frac{1}{1+\tau^{2}}
\int_0^{\infty}
w(\tau-u)(1+(\tau-u)^2)
\nn\\
\times
\exp\left\{
  -a \left[(1+\tau^2)u-\tau u^2+\frac{1}{3}u^3\right]
\right\}
\mathrm{d}u.
\end{eqnarray}
In order to evaluate this integral we will assume wave-vector
independence of the random turbulent noise $w(K(\tau))\approx
\mathrm{const}.$ For the real physics of accretion disks the bare
viscosity is evanescent and we have to analyze the above integral
on the $u$ variable for very small values of $a\ll1.$ This means
that for values of $|\tau|$ of the order of $1$ we have to take
into account very large values of the $u\gg1$ variable. For large
$u$ in the integrant we have to make the approximations
\begin{eqnarray}
1+(\tau-u)^2&\approx& u^2,\\
(1+\tau^2)u-\tau u^2+\frac{1}{3}u^3&\approx&\frac{1}{3}u^3,
\end{eqnarray}
and for the integral in the above \Eqref{BashSolution}
we derive according to \Eqref{pribl.br.chast.} the $\tau$-independent evaluation
\begin{eqnarray}
\label{br.paralelalfoni}
\frac{n_0}{w}
&\approx&\frac{1}{a}
%
\end{eqnarray}
%
%
%
In such a way for the low dissipation limit we derive an explicit
expression for the number of ``alfvons'' propagating along the
magnetic field
\be
\label{br.alfonizatau=0}
n_0\approx
\frac{Aw}{(\nu_k + \nu_m)k_z^2}
 \ee
The last term is expressed via the physical variables of the initial
problem and this is the central result of our analysis of the
kinetics of slow magnetosonic (Alfv\'{e}n) waves. Later, we will
perform statistical averaging of the dissipated power, but, before
that, in the next section, we will analyze in short the influence of
weak turbulence on the MHD equations.

\section{White noise turbulence}
\label{sec:WhiteNoise}
%
\subsection{Langevin-Burgers noise for Alfv\'{e}n waves}

In Langevin-Burgers approach to the turbulence the random noise is
described as a white one for which the following correlation may be used
\be \label{DispersiqNaBqlShum} \langle
\mathbf{F}(t_1,\mathbf{r}_1)\mathbf{F}(t_2,\mathbf{r}_2)\rangle
=\tilde\Gamma{\rho}^2\delta(t_1-t_2)\delta(\mathbf{r}_1-\mathbf{r}_2)\openone,
\ee
The dimension of $\tilde\Gamma$ shows that this correlation is
convenient not for forces, but rather for accelerations
\be
\left[\tilde\Gamma\right]={(\mbox{acceleration})}^2\times(\mbox{volume})\times(\mbox{time}).
\ee
We make a Fourier transform of the acceleration
\begin{eqnarray}
\mathbf{F}(t,\mathbf{r}) &=& \sum_{\mathbf{k}} \mathbf{F}_{k}(t)
                \exp (i\mathbf{k}\cdot\mathbf{r}),\\
\mathbf{F}_k(t) &=& \int \mathbf{F}(t,\mathbf{r})
                \exp(-i\mathbf{k}\cdot\mathbf{r})
\frac{\mathrm{d}^3x}{\mathcal{V}}
\end{eqnarray}
for which we apply periodic boundary conditions
\be
\mathbf{F}(t,x+L)=\mathbf{F}(t,x)
\ee
where $L$ is the characteristic length of the system
$\mathcal{V}=L^3.$ This sets the following restrictions for the wave
vectors
\be
(k_x,\, k_y,\, k_z,)= \frac{2\pi}{L}(n_x,\, n_y,\, n_z),
\ee
where the numbers can take only integer values
\be
n_x,\, n_y,\, n_z=0,\,\pm 1, \, \pm 2, \, \pm 3, \, \dots
\ee
In this case the difference between two wave vectors is $\Delta
k=2\pi/L$ and for great lengths the sum turns into an integral
\be
\frac{1}{\mathcal{V}}\sum_k=\int\frac{\mathrm{d}^3k}{(2\pi)^3}.
\ee
The correlation for the coefficients in the Fourier series is also a
white noise
\be \langle \mathbf{F}_p^*(t_1)\mathbf{F}_{k}(t_2)\rangle
=\openone\frac{\tilde\Gamma{\rho}^2}{\mathcal{V}}\delta(t_1-t_2)\delta_{pk},
\ee
where $\delta_{pk}$ is the symbol of Kronecker. Taking into account
the upper relation for the $x$ component of the acceleration we have
\be \langle F_{k,x}^*(t_1) F_{k,x}(t_2)\rangle
=\frac{\tilde\Gamma{\rho}^2}{\mathcal{V}}\delta(t_1-t_2). \ee
In our work we are interested in the presence of a white noise in
the MHD system (\ref{upsi=})~-~(\ref{bex=}) causing the primary
birth of the Alfv\'{e}n waves. That is why we want to know the
correlation for the dimensionless density of the external force
\be
\langle f^*(\tau_1) f(\tau_2)\rangle
=\Gamma \delta (\tau_1-\tau_2),
\ee
where
\be \label{Gamma_plane} f(\tau)\equiv\frac{F_{k,x}(t)}{\rho A
V_A},\quad \Gamma=\frac{\tilde\Gamma}{V_A^2A\mathcal{V}} \ee
and the parameter $\Gamma$ is dimensionless.

This dimensionless random force has to be taken into account as an
external force in the right hand side of \Eqref{deu_x} which now
takes the form
\be \label{deu_xNoise}
 \mathrm{d}_{\tau} \upsilon_x
  = \alpha(\tau)\, \upsilon_x - K_z b_x + f(\tau)
\ee
and analogously \Eqref{starteq} reads
\be \label{starteqNoise}
\ddot{\upsilon}
=\alpha(\tau)\dot{\upsilon} - K_z^2 \upsilon +\dot{f}(\tau).
\ee
Introducing the dimensionless displacement
\be
\label{displacement}
x(\tau)=\int_{-\infty}^{\tau}\upsilon(\tau')\mathrm{d}\tau'
\ee
we obtain
\be \label{starteqNoise}
\ddot{x}
=-\gamma_s(\tau)\dot{x} - K_z^2 x + f(\tau).
\ee
In such a way we arrived at the necessity to analyze the behavior of
classical harmonic oscillator with time-dependent friction and a
white noise as an external random force.

\subsection{General theorem for stochastic heating and income
term in the effective Boltzmann equation}
\label{GeneralTheorem}

The special cases considered in Appendix~\ref{app:stochHeating} for
a Brownian particle Eq.~(\ref{moshtnost}), an oscillator without
friction Eq.~(\ref{oscMoshnost}) and a free particle
Eq.~(\ref{freeMoshnost}) give us a hint that there is a general
formula for a white-noise stochastic heating. Let us apply this
common result to the effective oscillator equation
Eq.~(\ref{starteqNoise}) for the displacement of the plasma $x.$ In
case of negligible $\gamma_s$ Eq.~(\ref{oscMoshnost}) implies
\be
\mathrm{d}_\tau \left\langle\frac{1}{2}\dot x^2 + \frac{1}{2}K_z^2 x^2\right\rangle = \frac{\Gamma}{2},
\ee
where the dimensionless Burgers parameter $\Gamma$ is defined in
Eq.~(\ref{Gamma_plane}). At slow heating the virial theorem
$\left\langle \frac{1}{2}\dot x^2\right\rangle = \left\langle
\frac{1}{2}K_z^2 x^2\right\rangle$ gives that
\be
\mathrm{d}_\tau \left\langle\dot x^2\right\rangle = \frac{\Gamma}{2}.
\ee
According to the determination Eq.~(\ref{displacement}) $\dot x =
\upsilon$ and Eq.~(\ref{n=sredno(v^2)}) reads $n(\tau) =
\left\langle \dot x^2\right\rangle.$ In this way, in case of zero
friction $\gamma = 0,$ we obtain
\be
\label{www}
\mathrm{d}_\tau n(\tau) = w\equiv \frac{\Gamma}{2} = \frac{\tilde\Gamma}{2AV_A^2\mathcal{V}},
\ee
which is the source term in the right hand side of the Boltzmann
equation \Eqref{Boltzmanneq.}. The turbulent random forces generate
magnetosonic waves. Figuratively we may say that ``alfvons'' are
born from the \textit{turbulent sea foam}. The other terms in the
master equation \Eqref{Boltzmanneq.} describe the dissipative decay
with rate $\gamma_\nu$ and the ``lasing'' at $\gamma_s < 0.$

\section{Perspectives and conclusions}
\label{sec:concl}
%
The discussed mechanism requires simultaneous presence of both
magnetic field and turbulence -- ingredients which are pointed out
practically in all present works on the problem. As we have analyzed
MHD waves in an incompressible fluid the gas pressure turned out to
be irrelevant to the shear stress. Therefore we consider that in the
Shakura-Sunyaev phenomenology the gas pressure $p$ should be
replaced with the magnetic one $p_{_B}$. Thus for the momentum
transport we have
\be
\label{Debye}
\sigma_{\varphi r} = \alpha_{\omega} p_{_B}\;\mbox{  for disk},\quad
\sigma_{zr} = \alpha_{\omega} p_{_B}\;\mbox{  for column},
\ee
which is a rather small correction to the Shakura-Sunyaev
hypotheses.\cite{Shakura:72,Shakura-Sunyaev:73} The detailed
calculation of the dimensionless parameter $\alpha_{\omega}$
requires complete analysis of the kinetics for all modes in the
shear flow of the magnetized plasma. This is undoubtedly a very
complicated task and it is worth making a qualitative assessment for
the origin of the effective viscosity in a plasma with evanescent
initial one. The most important characteristic of the shear flow in
a magnetized plasma is the ``lasing'' phase when the wave draws
energy from the shear flow. Without dissipation this increment
reaches gigantic scales.\cite{Chagelishvili:93}

Complete analysis of the system, of course, requires detailed
numerical integration. Despite of this we can give a qualitative
explanation that the plasma heating and the arising of a huge
effective viscosity is caused by the ``laser'' amplification of the
convective instability. This is a rather universal mechanism playing
a significant role in our solar system in the diminution of the
angular momentum of our sun, but also it is the mechanism
responsible for the gigantic energy produced in quasars and AGNs. In
this way we offer a qualitative answer to the question why do the
most powerful sources of light in the universe shine - namely,
because of the instability of the slow magnetosonic waves in the
shear flow. We wish to point out that our theory for heating of
accretion disks is a conventional one, which does not require
additional hypotheses, but only numerical calculations. The energy
is produced in the bulk of the disk but finally emitted to its
sides.

The analyzed geometry $\mathbf{B} = (0, 0, B_0)$ corresponds, for
example, to an accretion column above the magnetic poles of a
neutron star. However, we suppose that the convective instability of
the slow magnetosonic waves is a general property of the magnetized
plasma with a shear flow. We believe that an arbitrary orientation
between the shear flow and the magnetic field would lead only to a
dimensionless multiplier of the order of unity in our final result.

The used Langevin-Burgers' approach is a first approximation for a
self-consistent treatment of the turbulence. We can modify it
including, for example, a wave-vector dependence of the noise
functions $\Gamma(k)$ in order to simulate the spectral density of
the velocity pulsations. Some parameter related to the spectral
density of the turbulence will indispensably participate in the
final result for the stress tensor. Moreover we have to include the
self-sustained turbulence in a realistic global model for the
accretion disks or accretion streams. In this direction we are not
giving a pret-a-porter prescription. The purpose of our work is only
to give an idea how the turbulence, magnetic field, and shear flow
play together in the most powerful sources of light in the universe
(for the mechanism of this shining we do not know more than the
girlfriend of H.~Bete on the energy source of stars, cf. the story
told by Feynman in his famous lectures). Simultaneously we can
observe traces of a big viscosity of the protoplanetary disk in the
angular momentum distribution in our solar system.\cite{Gurevich:78}
We just wish to rise the corner of the curtain and see the regally
play by: 1) the turbulence 2) magnetic field and 3) shear flow
creating the most efficient engine in the universe.

The work of this engine is related to the kinetics of the spectral
density of slow magnetosonic waves in a magnetized turbulent plasma
with a shear flow. The heating, effective viscosity and stress
tensors are statistical consequences of this spectral density. We
advocate that the ``lasing'' of ``alfvons'' is a key detail in the
accretion of many compact astrophysical objects in order to observe
the universe in its present form.

The last problem which we wish to speculate on is the applicability
of the Burgers' approach to magnetohydrodynamics. According to the
Lighthill theory\cite{Lighthill:52} the intensity of the emitted
sound is proportional to the ratio of the velocity pulsation and the
sound velocity to a high power $(v/v_s)^{5},$ see also
Ref.~[\onlinecite{L.VI}],~sec.~75. Qualitative considerations give
that the velocity of sound $v_s$ has to be substituted by the
Alfv\'{e}n velocity $V_A\ll v_s$ for small magnetic fields
$p_{_B}\ll p$. This leads to the conclusion that the transformation
of the turbulent energy to Alfv\'{e}n waves can be very effective
and those energies could be comparable. This qualitative property
was pointed out for the physics of solar plasma long time ago, see
for example Refs.~[\onlinecite{Schwarzschild:48,Parker:64}]. In such
a way in order to model the influence of the turbulence on MHD waves
we have to use high values of the noise intensity $\Gamma.$ This
parameter can be determined in such a way that the spectral density
of the waves' energy becomes equal to the spectral density of the
turbulent energy at a given cut-off wave-vector $k_c.$ For
qualitative computer simulations in the inertial regime we can use
these typical cut-off values as initial conditions and solve the MHD
equations without the stochastic force. This could be useful for a
numerical calculation of the spectral density of Alfv\'{e}n waves,
i.e. ``momentum'' distribution of alfvons in our conditional
terminology.

We conclude that it is necessary to revise the theoretical models of
disk accretion and we might expect appearance of a new direction in
the theoretical astrophysics, incorporating the methods of
statistical physics as a key detail in global magnetohydrodynamic
models for formation of compact astrophysical objects.

\acknowledgements The authors are thankful to Prof.~I.~Zhelyazkov
for the critical reading of the manuscript. This work was partially
supported by Bulgarian NSF grant VU-205-2006.

\bibliographystyle{apsrev}

\bibliography{AccretingDisks}

\begin{appendix}

\section{Matrix presentation for Lagrange--Euler transformations}
\label{app:matrixpresent}
 The transition between Lagrange and Eulerian coordinates
for the space- and wave-vectors is essential for the present paper.
For convenience of the reader we are giving these relations in a
transparent self-explainable matrix form:
\be \langle\mathbf{\tilde{k}}|\mathbf{\tilde{r}}\rangle
=\langle\mathbf{k}|\mathbf{r}\rangle,\qquad
\langle\mathbf{\tilde{k}}|=
\left(\begin{matrix}\tilde{k}_x&\tilde{k}_y&\tilde{k}_z\end{matrix}\right),
\ee
\be \langle\mathbf{\tilde{k}}|=
\left(\begin{matrix}\tilde{k}_x&\tilde{k}_y&\tilde{k}_z\end{matrix}\right)
=\left(\begin{matrix}k_x&k_y&k_z\end{matrix}\right) \left(
\begin{matrix}
1&0&0\\
0&1&0\\
\tau&0&1
\end{matrix}
\right), \ee
\be |\mathbf{\tilde{r}}\rangle
=\left(\begin{matrix}\tilde{x}\\\tilde{y}\\\tilde{z}\end{matrix}\right)
=\left(
\begin{matrix}
1&0&0\\
0&1&0\\
-\tau&0&1
\end{matrix}
\right) \left(\begin{matrix}x\\y\\z\end{matrix}\right), \quad
|\mathbf{r}\rangle
=\left(\begin{matrix}x\\y\\z\end{matrix}\right), \ee
\be |\mathbf{k}(t)\rangle=
\left(\begin{matrix}k_x\\k_y\\k_z\end{matrix}\right)
=\left(\begin{matrix}1&0&-\tau\\0&1&0\\0&0&1\end{matrix}\right)
\left(\begin{matrix}\tilde{k}_x\\\tilde{k}_y\\\tilde{k}_z\end{matrix}\right),
\ee
\be \left(
\begin{matrix}
1&0&0\\
0&1&0\\
-\tau&0&1
\end{matrix}
\right) \left(
\begin{matrix}
1&0&0\\
0&1&0\\
\tau&0&1
\end{matrix}
\right) =\openone, \ee
\be \left(
\begin{matrix}
1&0&-\tau\\
0&1&0\\
0&0&1
\end{matrix}
\right) \left(
\begin{matrix}
1&0&\tau\\
0&1&0\\
0&0&1
\end{matrix}
\right) =\openone. \ee

\section{Dynamics equations}
\label{app:dyneqs}

Let us consider the movement of incompressible magnetized
homogeneous plasma. The time evolution is given by the
system\cite{L.VIII}
\begin{align}
\label{NSeq.} \rho(\partial_t \mathbf{v} + \mathbf{v}\! \cdot\!
\nabla \mathbf{v})& = - \nabla p - \frac{1}{\mu_0}(\mathbf{B}
\times \texttt{rot}\, \mathbf{B}) + \eta \Delta \mathbf{v}
+ \mathbf{F},\\
\label{deB} \partial_t \mathbf{B} + (\mathbf{v}\! \cdot\!\nabla
\mathbf{B}) & = (\mathbf{B}\! \cdot \!\nabla)\mathbf{v} +
\frac{1}{\mu_0 \sigma}\Delta \mathbf{B}.
\end{align}
We presume small variations from the stable laminar current ($
\mathbf{v} = \mathbf{U}_0 + \mathbf{u}\, , \, \mathbf{B} =
\mathbf{B}_0 + \mathbf{B}^{'} ,\, \, p = p_0 + p^{'}$). Then, using
the assumptions for the direction of the external magnetic field and
the shear flow velocity ($\mathbf{B}_0 = B_0 \mathbf{e}_z,\,
\mathbf{U}_0 = Ax \mathbf{e}_z $), we find
\begin{align}
\nn (\mathbf{v} \cdot \nabla) \mathbf{v} = (\mathbf{U}_0 +
\mathbf{u})\cdot\nabla(\mathbf{U}_0 + \mathbf{u}) \\
\nn = (\mathbf{U}_0\cdot\nabla)\mathbf{U}_0 +
(\mathbf{u}\cdot\nabla)\mathbf{U}_0
+ (\mathbf{U}_0\cdot\nabla)\mathbf{u} \\
 = Au_x \mathbf{e}_z + Ax\partial_z\mathbf{u}.
\end{align}
Analogously, for the mixed multipliers we obtain
\begin{align}
\nn (\mathbf{v} \cdot \nabla) \mathbf{B} = (\mathbf{U}_0 +
\mathbf{u})\cdot \nabla(\mathbf{B}_0 +  \mathbf{B}^{'}) \\
\nn = (\mathbf{U}_0\cdot\nabla) \mathbf{B}_0 + (\mathbf{u}\cdot
\nabla)
\mathbf{B}_0 + (\mathbf{U}_0\cdot \nabla) \mathbf{B}^{'} \\
 =  Ax\partial_z\mathbf{B}^{'},
 \end{align}
and
\begin{align}
\nn (\mathbf{B} \cdot \nabla) \mathbf{v} = (\mathbf{B}_0 +
\mathbf{B}^{'})\cdot \nabla (\mathbf{U}_0 +
\mathbf{u}) \\
\nn = (\mathbf{B}_0\cdot\nabla)\mathbf{U}_0 +
(\mathbf{B}^{'}\cdot\nabla)\mathbf{U}_0 + (\mathbf{B}_0\cdot\nabla)\mathbf{u} \\
 = B_0\partial_z\mathbf{u} + AB^{'}_x \mathbf{e}_z, \\
\nabla p = \nabla p^{'}\, , \,\, \Delta \mathbf{B} =
\Delta\mathbf{B}^{'}\, , \,\,\Delta\bf{v} = \Delta\mathbf{u}.
\end{align}
The time derivative of both variables is expressed by their varying
components
\begin{align}
\partial_t \mathbf{v} =
\partial_t \mathbf{u} \, , \, \,\partial_t \mathbf{B} = \partial_t
\mathbf{B}^{'}.
\end{align}
Having in mind the above mentioned assumptions the vector product of
the magnetic field and its rotation may be rewritten in the form
\begin{align}
\nn \mathbf{B} \times \mathrm{rot}\, \mathbf{B} = \mathbf{B}_0
\times
\mathrm{rot}\, \mathbf{B}^{'} \\
\nn  =\left(
\begin{matrix}
(\mathbf{B}_{0})_y(\mathrm{rot}\mathbf{B}^{'})_z -
(\mathbf{B}_{0})_z(\mathrm{rot}\mathbf{B}^{'})_y \\
(\mathbf{B}_{0})_z(\mathrm{rot}\mathbf{B}^{'})_x -
(\mathbf{B}_{0})_x(\mathrm{rot}\mathbf{B}^{'})_z \\
(\mathbf{B}_{0})_x(\mathrm{rot}\mathbf{B}^{'})_y -
(\mathbf{B}_{0})_y(\mathrm{rot}\mathbf{B}^{'})_x \\
\end{matrix}
\right) \\
= \left(
\begin{matrix}
- B_{0}(\mathrm{rot}\mathbf{B}^{'})_y \\
 B_{0}(\mathrm{rot}\mathbf{B}^{'})_x \\
  0
\end{matrix}
\right) =
 B_0 \! \! \left(
\begin{matrix}
\partial_{x} B^{'}_{z} - \partial_{z} B^{'}_{x} \\
\partial_{y} B^{'}_z  - \partial_{z} B^{'}_y \\
0
\end{matrix}
\right).
\end{align}
Consequently, Eqs.~(\ref{NSeq.}, \ref{deB}) turn into
\begin{align}
\nn \rho(\partial_t\mathbf{u} + Ax\partial_z \mathbf{u} +
 Au_x\mathbf{e}_z) = -\nabla p^{'} \\
+ \,\,\eta \Delta \mathbf{u} + \frac{B_0}{\mu_0} \left(
\begin{matrix}
\partial_z B^{'}_x - \partial_x B^{'}_z \\
\partial_z B^{'}_y - \partial_y B^{'}_z\\
0 \end{matrix} \right)\!+ \mathbf{F}, \\
\partial_t \mathbf{B}^{'} +
Ax\partial_z\mathbf{B}^{'} = B_0\partial_z \mathbf{u} +
AB^{'}_x\mathbf{e}_z + \frac{1}{\mu \sigma}\Delta\mathbf{B}^{'}.
\end{align}
%


%

\section{Derivation of the MHD set of equations with random noise}
\label{app:MHDsyst}
Substituting a plane anzatz waves in the linearized
MHD Eqs.~(\ref{Euler})
\begin{eqnarray}
\label{PlaneWaveAnzatz}
\mathbf{u}(t,\mathbf{r})
&=&-iV_A \, \vec{\upsilon}(\tau)\exp(i\mathbf{k}(t)\cdot\mathbf{r}),\\
\mathbf{B}^{\prime}(t,\mathbf{r})
&=& B_0 \, \mathbf{b}(\tau)\exp(i\mathbf{k}(t)\cdot\mathbf{r}),\\
p(t,\mathbf{r})
&=& \rho V_A^2 \,P(\tau) \exp(i\mathbf{k}(t)\cdot\mathbf{r}),\\
\mathbf{F}(t,\mathbf{r})
&=& -i \rho A V_A \, \mathbf{f}^{\prime}
(\tau)\exp(i\mathbf{k}(t)\cdot\mathbf{r}),
\end{eqnarray}
where according to Eqs.~(\ref{div_B}), (\ref{div_u}), and
(\ref{kEuler})
\be
\label{divZero}
\mathbf{K}\cdot\mathbf{b}=0,\qquad \mathbf{K}\cdot\vec{\upsilon}=0,
\ee
\begin{equation}
\label{K_tau}
\mathbf{k}(t)=\frac{A}{V_A}\mathbf{K}(\tau),\;
\mathbf{K}(\tau)=
\left(
\begin{matrix}
-\tau K_z\\K_y\\K_z
\end{matrix}
\right),\;
\dot{\mathbf{K}}=K_z\mathbf{e}_z,
\end{equation}
we obtain a system of ordinary differential equations
\begin{eqnarray}
\label{upsilon_system}
\dot{\vec{\upsilon}}&=&
-\upsilon_x \mathbf{e}_z+\mathbf{K} P \nn\\
&&-\left(
\begin{matrix}
K_zb_x-K_xb_z\\
K_zb_y-K_yb_z\\
0
\end{matrix}
\right)
-\nu_k'K^2\vec{\upsilon}+\mathbf{f}',
\\
\label{b_prime} \dot{\mathbf{b}} &=&
K_z\vec{\upsilon}+b_x\mathbf{e}_z-\nu_m'K^2\mathbf{b} .
\end{eqnarray}
Here $\tau=At,$ the dot operation stands for $\tau$-differentiation
$\mathrm{d}_\tau$ and we have introduced dimensionless notations for
both kinematic and magnetic viscosities
\be
\nu_k'=\frac{A}{V_A^2}\nu_k\qquad \nu_m'=\frac{A}{V_A^2}\nu_m.
\ee
All dimensionless variables in the upper system are expressed in
specific units: for velocity $V_A,$ pressure $\rho V_A^2,$ magnetic
field $B_0,$ time $1/A,$ wave-vector $A/V_A,$ kinematic viscosity
$V_A^2/A,$ acceleration $AV_A,$ density of force $\rho A V_A,$ and
length $V_A/A.$

Time differentiation of Eq.~(\ref{divZero}) gives
\be
K_x\dot{\upsilon}_x+K_y\dot{\upsilon}_y+K_z\dot{\upsilon}_z
-K_z\upsilon_x=0.
\ee
The substitution here of $\dot{\vec{\upsilon}}$ from
Eq.~(\ref{upsilon_system}) gives for the pressure
\begin{equation}
\label{pressureP}
P=-b_z+2\frac{K_z}{K^2}\upsilon_x-\frac{1}{K^2}(K_x f_x'+K_y f_y'+K_z f_z'),
\end{equation}
and back substitution of the dimensionless pressure in the x- and
y-components of Eq.~(\ref{upsilon_system}) gives the final equations
for the acceleration
\begin{eqnarray}
\label{dot_upsilon}
\dot{\upsilon}_x&=&2\frac{K_zK_x}{K^2}{\upsilon_x}-K_zb_x-\nu'K^2\upsilon_x
\nn\\
&&+f_x'-\frac{K_x}{K^2}(K_x f_x'+K_y f_y'+K_z f_z'),\\
\dot{\upsilon_y}&=&2\frac{K_zK_y}{K^2}{\upsilon_x}-K_zb_y-\nu'K^2\upsilon_y
\nn\\
&&+f_y'-\frac{K_y}{K^2}(K_x f_x'+K_y f_y'+K_z f_z').
\end{eqnarray}
For the magnetic field we take the x- and y-components from
Eq.~(\ref{b_prime})
\begin{eqnarray}
\label{dot_b}
\dot{b}_x&=& K_z\upsilon_x-\nu_m'K^2b_x,\\
\dot{b}_y&=& K_z\upsilon_y-\nu_m'K^2b_y.
\end{eqnarray}
The corresponding equations for the z-components
\begin{eqnarray}
\dot{\upsilon}_z&=&\left(-1 + 2\frac{K_z^2}{K^2}\right)
   \upsilon_x - K_z b_z-\nu_k'K^2\upsilon_z\nn\\
&&+f_z'-\frac{K_z}{K^2}(K_x f_x'+K_y f_y'+K_z f_z'),\\
\dot{b_z}&=&K_z\upsilon_z+b_x-\nu_m'K^2b_z
\end{eqnarray}
are not necessary to be solved;
the solutions are given by Eqs.~(\ref{divZero})
\begin{eqnarray}
\upsilon_z&=&-(K_x\upsilon_x+K_y\upsilon_y)/K_z, \\
b_z&=&-(K_x b_x+K_y b_y)/K_z.
\end{eqnarray}

\subsection{2D restriction with $K_y=0$}

In the special case of $K_y=0$ we have separation of variables and
the y-component of the acceleration is
\begin{equation}
\dot{\upsilon_y}=-K_zb_y-\nu'K^2\upsilon_y+f_y',
\end{equation}
where for the Langevin force we suppose a white noise correlator
\begin{equation}
\langle \mathbf{f}'(\tau_1) \mathbf{f}'(\tau_2)\rangle
= \openone \Gamma \delta(\tau_1-\tau_2),\quad
\Gamma= \frac{\tilde{\Gamma}}{V_A^2A\cal{V}}.
\end{equation}
In such a way we derive Eqs.~(\ref{upsi=}, \ref{upsiy=}).

In the present work we will use the $K_y=0$ case for a model
evaluation of the statistical properties of Alfv\'en waves.

\subsection{Test example of short wavelengths and small attenuation}

As a test example let us finally analyze the textbook's case of a
negligible viscosity and big enough wave-vector $K_z.$ The systems
for $\upsilon_x,$ $b_x$ and $\upsilon_y,$ $b_y$ have approximate
time dependent solution $\propto \exp\left( -i|K_z|\tau\right).$
This means that both $x$- and $y$-polarized modes have dispersion of
Alfv\'{e}n waves $\omega \approx A |K_z|$ or more precisely
\be
 \omega(\mathbf{k})=\omega_A
 = V_A|k_z|- \frac{i}{2}\nu k^2,\quad
\omega_A \gg A, \; \nu k^2. \ee
This imaginary term may be obtained as the first Newton correction
in the method described above (see
Eq.~(\ref{firstNewtonIteration})).
 Let us mention that the attenuation given by the small imaginary
part of the frequency is an isotropic function of the wave vector,
cf. Ref.~[\onlinecite{L.VIII}], sec.~69, problem.

We analyze an inviscid approximation and that is why for both MHD
branches the oscillations of the velocity $\mathbf{u}$ are
transversal to the wave-vector $\mathbf{k}.$ For the y-mode the
velocity oscillation $\mathbf{u}$ is perpendicular to the
$\mathbf{k}$-$\mathbf{B}_0$-plane; this is the true Alfv\'{e}n wave.
For the x-mode, which we are analyzing in the current paper, the
displacement of the fluid and the magnetic force lines lies in the
$\mathbf{k}$-$\mathbf{B}_0$-plane. For finite compressibility the
x-mode is hybridized with the sound, that is why it is often called
slow magnetosonic wave even if at low magnetic fields $p_{_B}\ll p$
its dispersion is given by the Alfv\'{e}n waves' dispersion
$\omega_A\approx V_A| \mathbf{k} \cdot \mathbf{B}_0|/B_0.$

\section{Attenuation coefficient for Alfv\'{e}n waves}
\label{app:koefnazatihv}

In order to derive the dissipative function for Alfv\'{e}n waves
spreading in a highly conducting plasma, we have to know the
energy and its variation in time due to kinematic and magnetic
friction. In this section we are interested only in the average
dissipation, that is why it is convenient to work with averaged
(with respect to the wave's phase) quantities. The density of
electromagnetic energy is given by
\be \mathcal{E}_{\mathrm{em}} \approx \frac{B^2}{2\mu_0} , \ee
where we have used that for a highly conducting media the
electromagnetic energy $E^2/2\varepsilon_0\propto 1/\sigma^2$ is
negligible compared to the magnetic one.

We are interested in the wave part of both the magnetic and the
velocity fields, as the constant part does not play any role in the
dynamics of the system. Rewritten in terms of the dimensionless wave
components (see Eq.~(\ref{B_plane})) the expression for the energy
density looks like
\be \mathcal{E}_{\mathrm{em}}\approx b^2p_{_B}, \ee
where the magnetic pressure $p_{_B} = B_0^2/2\mu_0$ has already been
introduced in the text above (sec.~\ref{sec:Kinematics},
Eq.~(\ref{pressuremagnitno})). According to the equipartition
(Alfv\'en) theorem the Alfv\'{e}n waves kinetic energy density
equals the magnetic one
\be \mathcal{E}_{\mathrm{kin}} = \frac{1}{2}\rho V^2 = \upsilon^2
p_{_B}. \ee
Then having in mind that
\be \label{ravenstvoNaSrednite}
 \left\langle\upsilon_x^2\right\rangle = \left\langle
b_x^2\right\rangle,\quad \left\langle\upsilon_y^2\right\rangle =
\left\langle b_y^2\right\rangle \ee
for the averaged volume density of the total energy we obtain
\begin{align}
\mathcal{E}_{\mathrm{tot}}=2p_{_B}\left[(1+\tau^2)\left \langle
\upsilon_x^2\right \rangle + \left \langle\upsilon_y^2\right
\rangle\right].
\end{align}
For an estimation of the attenuation coefficient one also needs to
know the averaged with respect to the waves' period volume density
of the total power $\mathcal{Q}_{\mathrm{tot}},$ which is a sum of
the averaged kinetic and magnetic one
\be \label{srednaMo6tnost} \mathcal{Q}_{\mathrm{tot}}=\left
\langle \mathcal{Q}_{\mathrm{kin}}\right \rangle + \left \langle
\mathcal{Q}_{\mathrm{em}}\right \rangle. \ee
The kinetic power is simply the time derivative of the kinetic
energy and since we are interested in the wave component of the
velocity it is given by
\be \mathcal{Q}_{\mathrm{kin}} = \rho \mathbf{u}_{\tilde
k}(t)\cdot\mathrm{d}_{\tilde{t}}\mathbf{u}_{\tilde k}(t), \ee
where the separation of variables Eq.~(\ref{plane-wave}) and the
linearization Eqs.~(\ref{razdNaProm}, \ref{sd_lin_u}) have been
used.

Rewritten in the dimensionless variables Eq.~(\ref{bezrazmSkorost})
with the incompressibility condition Eq.~(\ref{pulnanesviv}) applied
this reads
\begin{align}
\nn \mathcal{Q}_{\mathrm{kin}} =
-2Ap_{_B}\vec{\upsilon}\cdot\mathrm{d}_{\tau}\vec{\upsilon}\\
\nn =\upsilon_x\left(1+\frac{K_x^2}{K_z^2}\right)\!
\left[\frac{K_xK_z}{K^2}\upsilon_x - K_zb_x - \nu_k^{\prime}K^2\upsilon_x + gf_x\right]\\
+\; \upsilon_y\left(1+\frac{K_y^2}{K_z^2}\right)\!
\left[\frac{K_yK_z}{K^2}\upsilon_x - K_zb_y -
\nu_k^{\prime}K^2\upsilon_y + hf_y\right].
\end{align}
When we take an average value of the upper expression only the
quadratic terms remain
\be \left\langle\upsilon_x\right\rangle = \left\langle
b_x\right\rangle = \left\langle\upsilon_y\right\rangle =
\left\langle b_y\right\rangle = 0, \ee
therefore for the kinetic part of the average power we have
\begin{align}
\nn\left\langle\mathcal{Q}_{\mathrm{kin}}\right\rangle = -2Ap_{_B}
\left[\left(1+\frac{K_x^2}{K_z^2}\right)
\left(\frac{K_xK_z}{K^2} - \nu_k^{\prime}K^2\right)\langle\upsilon_x^2\rangle\right.\\
\left.-
\left(1+\frac{K_y^2}{K_z^2}\right)\nu_k^{\prime}K^2\langle\upsilon_y^2\rangle\right].
\end{align}
If we use the relation Eq.~(\ref{ratio}) for the special case when
$K_y = 0$ this yields
\begin{align}
\label{srednaQ_kin} \nn\langle\mathcal{Q}_{\mathrm{kin}}\rangle =
2Ap_{_B}K_z^2\left({(1+\tau^2)}^2\nu_k^{\prime}\langle\upsilon_x^2\rangle
+ (1+\tau^2)\nu_k^{\prime}\langle\upsilon_y^2\rangle\right) \\
+ 2Ap_{_B}\tau\langle\upsilon_x^2\rangle.
\end{align}
Now we have to calculate the Ohmic part of the power in the
magnetostatic approximation Eq.~(\ref{Maxeqn})
\begin{align}
\nn \mathcal{Q}_{\mathrm{em}} = \mathbf{j}\cdot
\mathbf{E^{^\prime}} = \frac{j^2}{\sigma}
=\frac{{(\mathrm{rot}\,\mathbf{B})}^2}{\mu_0^2 \sigma}\\
=2p_{_B}\nu_m{(\mathrm{rot}\,\mathbf{b})}^2
=2Ap_{_B}\nu_m^{\prime}{(\mathbf{K}\times \mathbf{b})}^2,
\end{align}
where $\mathbf{E^\prime}$ is the effective electric field in the
Ohm's law Eq.~(\ref{Ohm2}), $\mathbf{b}$ is the dimensionless
magnetic field defined in Eq.~(\ref{bezrazmMagnP}) and
$\nu_m^\prime$ is the dimensionless magnetic viscosity given in
Eq.~(\ref{bezrazmMagnVisc}).

In the particular case when $K_y = 0$ the upper expression in
terms of the dimensionless time $\tau$ turns into
%
%
%
\begin{align}
\mathcal{Q}_{\mathrm{em}} = 2Ap_{_B}\nu_m^{\prime}K_z^2(1+\tau^2)
\left[(1+\tau^2)b_x^2 + b_y^2\right].
\end{align}

Therefore if we take into account Eqs.~(\ref{ravenstvoNaSrednite},
\ref{srednaMo6tnost}, \ref{srednaQ_kin}) and the equation above
for the average density of the total power dissipated in the fluid
we have
\begin{align}
\mathcal{Q}_{\mathrm{tot}}
=2Ap_{_B}\nu^{\prime}K_z^2(1+\tau^2)((1+\tau^2)\upsilon_x^2 +
\upsilon_y^2) + 2Ap_{_B}\tau\upsilon_x^2,
\end{align}
where $\nu$ is the total viscosity defined in
Eq.~(\ref{pulenMagnVisc}).

Now it can be easily shown that the attenuation coefficient takes
the following form
\be \overline\gamma =
\frac{\mathcal{Q}_{\mathrm{tot}}}{\mathcal{E}_{\mathrm{tot}}} =
A(1+\tau^2)\nu^{\prime}K_z^2 + \frac{A\tau}{1+\tau^2}. \ee
In this way we derived the approximative (in case of small
dissipation) equation Eq.~(\ref{decrement}).

\section{Stochastic heating}
\label{app:stochHeating}

\subsection{Kinetic equation for the kinetic energy of a Brownian particle}

The Langevin\cite{Langevin:08} approach for a treatment of
stochastic differential equations is practically unknown in
astrophysics; most of the actively working in this field people even
have not heard about Langevin-Burgers' approach to
turbulence\cite{Burgers:48}. That is why instead of referring to
textbooks far away from our current problem\cite{Gardiner:85} we
will give a pedestrian introduction of all the necessary basic
notions. In the framework of the used notations we will introduce
the necessary mathematics. Our first illustration will be the
diffusion of a Brownian particle. The theory is analogous to the
Nyquist\cite{Nyquist} theory for the thermal noise in electric
circuites.

The literature on \textit{``Burgers turbulence''} is really huge
http://google.com search gives 29, 000 items for 2007, or if we
introduce a google index function it will read: $\mathcal
G\left(\mbox{``Burgers turbulence''\!, 2007}\right)= 29,\;000.$
Analogously, for Langevin turbulence we have $\mathcal
G\left(\mbox{Langevin turbulence, 2007}\right) = 113,\,000.$ We will
mention only several reviews on this subject\cite{BurgersPapers}.

Our goal in this subsection is to derive an explicit and physically
grounded expression for the random noise in the Boltzmann equation
Eq.~(\ref{Boltzmanneq.}). We shall start our deduction by
consideration of the kinetic equation for a Brownian particle,
moving with velocity $\mathrm{v}$ under the influence of a random
force $\mathbf{\hat f}(t)$ across a given medium with a ``friction
coefficient'' $\tilde\gamma$
\be \label{kineqn} m \mathrm{d_t} \mathbf{v} = - m \tilde\gamma
\mathbf{v} + \mathbf{\hat f}(t). \ee
For the sake of simplicity from now on we shall examine the
propagation of the particle in $x$ direction, so that we shall be
interested only in the projections of the velocity and the external
force along this axis, which we shall designate with $v$ and $\hat
f$. The upper ordinary differential equation may be easily solved
with the help of the Euler method
\be v(t) = C(t)\exp(-\tilde\gamma t). \ee
We substitute the velocity from Eq.~(\ref{kineqn}) with the given
expression and obtain another ordinary differential equation with
separable variables for the unknown function $C(t)$
\be \exp(- \tilde\gamma t)\mathrm{d}_tC(t) = \frac{\hat f(t)}{m},
\ee
whose solution is
\be C(t) = v_0 + \int^{t}_{t_0} \frac{\hat f(t_1)}{m}
\exp(\tilde\gamma t_1)\mathrm{d}t_1. \ee
The constant of integration $v_0$ here stands for the initial
velocity. The explicit form of the velocity becomes
\be v(t) = \exp(-\tilde\gamma t)\left( v_0 + \int^{t}_{t_0}
\frac{\hat f(t_1)}{m} \exp(\tilde\gamma t_1)\mathrm{d}t_1 \right).
\ee
Now let us average the square of this velocity. For this purpose we
need to know the correlation of the noise. Following the classical
works by Langevin and Burgers\cite{Burgers:48,Langevin:08} we
suppose that the external force correlator has the simplest possible
form of a white noise
\be \left\langle \hat f(t_1)\hat f(t_2) \right\rangle = \hat\Gamma
\delta(t_1 - t_2). \ee
We take into consideration that the random force has zero mean value
$\left\langle \hat f(t) \right\rangle = 0$ and for the average of
the square of the velocity we obtain
\begin{eqnarray}
\nn\left\langle v^2(t) \right\rangle = \exp (-2\tilde\gamma t)  \\
\nn \times \left( v_0^2 + \int^{t}_{t_0}
\int^{t}_{t_0}\frac{\hat\Gamma \delta(t_1 - t_2)}{m^2}
\exp(\tilde\gamma (t_1 + t_2)) \mathrm{d}t_1\mathrm{d}t_2 \right) \\
\nn=\exp(-2\tilde\gamma t)\left( v_0^2 +
\!\int^{t}_{t_0}\frac{\hat\Gamma}{m^2} \exp(2\tilde\gamma t_1)
\mathrm{d}t_1
\right).\\
\end{eqnarray}
This noise averaging is the most important ingredient in the present
derivation. It reduces the mechanical problems to a statistical
problem solvable by the Boltzmann equation. In the next subsection
we apply this approach to a harmonic oscillator and later on to the
amplitude of magnetosonic waves.

We are now ready to calculate the mean kinetic energy of the
Brownian particle
\begin{eqnarray}
\nn\left\langle E_{\mathrm{kin}}(t) \right\rangle
= \frac{m}{2}\left\langle v^2(t) \right\rangle \\
\nn= \frac{m v_0^2}{2}\exp(-2\tilde\gamma t) +
\frac{\hat\Gamma}{2m}\exp(-2\tilde\gamma t) \int^{t}_{t_0}
\exp(2\tilde\gamma t_1)\mathrm{d}t_1 \\
=\frac{m v_0^2}{2}\exp(-2\tilde\gamma t) +
\frac{\hat\Gamma}{4m\tilde\gamma}\left\{1 - \exp(-2\tilde\gamma (t -
t_0))\right\}.
\end{eqnarray}
To rewrite this expression in a more convenient way we introduce the
initial kinetic energy $E_0 = E_{kin}(t_0)$ and the equilibrium one
$\overline{E} \equiv \hat\Gamma/4m\tilde\gamma.$ Thus the average
kinetic energy reads
\be \label{Ekin} \left\langle E_{\mathrm{kin}}(t) \right\rangle =
E_0 \exp \left(-\frac{t}{\tau_E}\right) + \overline{E} \left\{1 -
\exp\left(-\frac{(t - t_0)}{\tau_E}\right)\right\}, \ee
where $\tau_E \equiv 1/2\tilde\gamma$ corresponds to the relaxation
time in the atomic physics and $\hat\Gamma$ is the analogue to the
width of the Lorentz function. Time differentiation of
Eq.~(\ref{Ekin}) gives the kinetic equation for the kinetic energy
\begin{eqnarray}
\nn \mathrm{d}_t\left\langle E_{\mathrm{kin}}(t)\right\rangle =
-\frac{1}{\tau_{E}}\left( E_0\exp
\left(-\frac{t}{\tau_E}\right)\right.\\
\left. + \; \overline{E} \exp\left(-\frac{(t -
t_0)}{\tau_E}\right)\right).
\end{eqnarray}
In this way we came to the well-known Boltzmann equation for the
variation of the average kinetic energy
\be \frac{\mathrm{d}}{\mathrm{d}_t}\left\langle
E_{\mathrm{kin}}(t)\right\rangle = -\frac{1}{\tau_{E}}\left(
\left\langle E_{\mathrm{kin}}(t)\right\rangle - \overline{E}
\right). \ee
The first term here is responsible for the energy expenditure and
consequently is set by the dissipation function, whereas the second
stands for the energy income (i.e. a positive power), caused by the
effect of the random noise.

Now let us make a short analysis of the result. In the case of
equilibrium (after several relaxation times, $t \rightarrow \infty$)
the mean value of the kinetic energy equals half the absolute
temperature which we will designate with $\Theta$
\begin{align} \left\langle E_{\mathrm{kin}}(t)\right\rangle =
\frac{m}{2}\left\langle v^2\right\rangle \\
=\frac{1}{2}\Theta = \overline{E} =
\frac{\hat\Gamma}{4m\tilde\gamma}.
\end{align}
Hereby, we can deduce the connection between the dissipation
coefficient $\tilde\gamma$ and the fluctuations $\hat\Gamma$
\be \hat\Gamma = 2m\Theta \tilde\gamma, \ee
which is a special case of the fluctuation-dissipation theorem.

Our idea is to apply notions from the equilibrium statistics in the
non-equilibrium case, therefore we shall use $\Theta$ as a white
noise parameter in stead of $T$ (a non-equilibrium analogue to the
equilibrium temperature). In terms of the averaged power of energy
dissipation $Q,$ the fluctuation-dissipation theorem reads
\be \label{moshtnost} Q = \frac{\overline{E}}{\tau_E} = \tilde\gamma
\Theta = \frac{\hat\Gamma}{2m}. \ee

The first special case of the fluctuation dissipation theorem is the
relation between diffusion coefficient and the mobility of a
Brownian particle. We will remind some details. Let us consider the
motion of a Brownian particle as a diffusion, for which the role of
the concentration is played by the probability to find the particle
at a given place in the volume of the fluid. The diffusion equation
is
\be
\partial_t n(x,t) = D \Delta n(x,t),
\ee
where $n(x,t)$ stands for the concentration. The square of the
average distance, which the particle passes for an interval of
time $t$ is given by Ref.~[\onlinecite{L.VI}],~sec.~(60)
\be \left\langle x^2 \right\rangle = \int x^2 n(x,t)\mathrm{d}x =
2Dt. \ee
According to the Sutherland-Einstein
relation\cite{Sutherland:05,Einstein:05,Pais:82} the diffusion
coefficient $D$ is proportional to the mobility $\mu$ and the
temperature in the equilibrium statistical physics, which under
our assumption may be expressed as
\be D= \mu\Theta. \ee
The mobility may be derived from the equation of motion of the
particle, roaming under the influence of a time independent random
force $\mathbf{\hat f}$
\be m\mathrm{d}_t\mathbf{v} = -m\tilde\gamma \mathbf{v} +
\mathbf{\hat f} . \ee
In the stationary case the driving force has to balance the friction
\be v_\mathrm{dr} = \mu\hat f. \ee
Hereby for the mobility we obtain
\be \mu = \frac{1}{m\tilde\gamma} \ee
and the diffusion coefficient takes the form
\be D = \frac{\Theta}{m\tilde\gamma}. \ee
Our next step is to consider the equation of motion for a harmonic
oscillator under a random noise.

\subsection{Oscillator under a white noise}

In a self-consistent linearized approximation the problem for the
wave propagation is reduced to independent oscillator problems for
all wave vectors. That is why the Brownian motion of a harmonic
oscillator is a key detail of our statistical theory for the
spectral density of magnetosonic waves. We are starting with the
equation of motion for a harmonic oscillator with an external white
noise
\be m\ddot{\mathbf{x}} = - m \omega^2\mathbf{x} + \mathbf{\hat
f}(t). \ee
We assume classical oscillator (i.e. $[\mathbf{x},\mathbf{p}] = 0$).
It is convenient to examine the equation of motion in the phase
space. For that purpose we introduce complex variables
\begin{align}
\nn c = m\omega\mathbf{x} + i\mathbf{p}\\
c^* = m\omega\mathbf{x} - i\mathbf{p},
\end{align}
where $\mathbf{p} = m \dot{\mathbf{x}}$ is the particle momentum.

The energy of the oscillator (proportional to the number of
particles) can be expressed via these variables
\be \label{E=b+b}E = \frac{1}{2m}c^*c = \frac{p^2}{2m} +
\frac{1}{2}m\omega^2 x^2 \ee
We are interested in the explicit form of the mean power $Q =
\mathrm{d}_t \left\langle E(t) \right\rangle.$ That is why we want
to know the correlation between $c$ and $c^*$. In order to find it
first we have to determine their explicit forms as functions of
frequencies, random noises and time. The time derivative of $c$
looks like
\begin{align}
\nn \mathrm{d}_tc = m\omega \dot{\mathbf{x}} + i\dot{\mathbf{p}} =
\omega\mathbf{p} + i\left(
-m\omega^2\mathbf{x} + \mathbf{\hat f}(t)\right)\\
= -i\omega(i\mathbf{p} + m\omega\mathbf{x}) = -i\omega c(t) +
i\mathbf{\hat f}(t).
\end{align}
For this ordinary differential equation we seek a solution in the
form of harmonic oscillations
\begin{align} \nn c(t) = \left( C_0 + i
\int^t_{t_0} \hat f(t_1) \exp(i\omega
t_1)\mathrm{d}{t_1}\right)\exp(-i\omega t)\\
c^*(t) = \left( C_0 - i \int^t_{t_0}\hat f(t_2) \exp(-i\omega
t_2)\mathrm{d}{t_2}\right)\exp(-i\omega t).
\end{align}
Thus for the correlator we have
\begin{align}
\nn \left\langle c^*(t)c(t)\right\rangle = {|C_0|}^2 \\
 \label{cc*} + \int^t_{t_0}\int^t_{t_0} \left\langle
\hat f(t_1)\hat f(t_2) \right\rangle \exp(i\omega(t_1 -
t_2))\mathrm{d}{t_1}\mathrm{d}{t_2}.
\end{align}
We consider a white noise for which $\left\langle \hat f(t_1)\hat
f(t_2)\right\rangle = \hat\Gamma \delta(t_1~-~t_2).$ Then
Eq.~(\ref{cc*}) turns into
\be \left\langle c^*(t)c(t)\right\rangle = {|C_0|}^2 + \hat\Gamma
(t-t_0). \ee
This averaging is one of the most important details of the present
theory -- it reduces the wave problem to a statistical one solvable
by the Boltzmann equation.

Finally for the averaged energy of the oscillator under the random
noise we have
\be \left\langle E(t) \right\rangle = E_0 +
\frac{\hat\Gamma}{2m}(t-t_0), \ee
where the initial energy is $E_0 = \frac{1}{2m}{|C_0|}^2.$ The
produced power is the same as in Eq.~(\ref{moshtnost})
\be \label{oscMoshnost} Q = \frac{\hat\Gamma}{2m}. \ee
Here we wish to mention that analogous scenario of linear dependence
of the energy of ocean waves driven by turbulent fluctuations of the
pressure was proposed by Jeffries, Fillips, Feynman and Hibbs. This
model, however, could be applicable only before the creation of a
system of parallel vortices which are the real intermediary between
the wind and waves [\onlinecite{TH1,TH2}]; see also the references
therein.

The nature of the friction force and the external potential is
irrelevant to this result because by definition the white noise is a
very fast process. In order to illustrate this general theorem we
will analyze in the next section the averaged power of a free
particle under a white noise.

\subsection{Stochastic heating of a free particle}
This phenomenon is analogous to the Fermi model for acceleration of
cosmic particles. The equation of motion for a free particle reads
\be m \dot{v} = \hat f(t), \ee
where $\hat f(t)$ stands for the random (under our consideration
white) noise. Hereby, after integration, for the velocity vector we
find
\be v(t) = \int_{t_0}^{t} \frac{\hat f(t)}{m}\mathrm{d}t_1 + v_0,
\ee
where $v_0$ is the initial velocity. Using this explicit expression
the average kinetic energy reads as
\begin{align}
\nn \left\langle E(t)\right\rangle = \frac{m}{2}\left\langle v^2(t)\right\rangle \\
= \frac{1}{2m}\int^{t}_{t_0}\int^{t}_{t_0} \left\langle \hat
f_1(t)\hat f_2(2)\right\rangle \mathrm{d}t_1\mathrm{d}t_2 +
\frac{m}{2}\left\langle v_0^2 \right\rangle.
\end{align}
Here we apply the correlation of the white noise $\left\langle \hat
f_1(t)\hat f_2(2)\right\rangle = \hat\Gamma \delta(t_1~-~t_2)$ and
integrate with respect to time. Then the average kinetic energy
becomes
\be \left\langle E(t)\right\rangle = E_0 + \frac{\hat\Gamma}{2m}(t -
t_0). \label{erqvx} \ee
Finally a time differentiation gives the power of the white noise
acting on the free particle
\be \label{freeMoshnost} Q = \mathrm{d}_t\left\langle E(t)
\right\rangle = \frac{\hat\Gamma}{2m}. \ee
Again this is the same result as in Eqs.~(\ref{moshtnost}) and
(\ref{oscMoshnost}). This constant power is applicable if the
typical relaxation times of the particle are much bigger than the
characteristic times of the noise and we can approximate it as a
white one. When we apply this result to the averaged amplitude of
the magnetosonic waves this constant power gives the constant income
term $w$ in the Boltzmann equation for the spectral density of the
MHD waves.

\section{Numerical Analysis}
\label{app:numerical}

\subsection{Pure hydrodynamics}
\label{app:purehydrodyn}

Our first step in the numerical analysis is to investigate the
influence of the shear flow for amplification of initial
perturbation in framework of pure hydrodynamics at zero magnetic
field $\mathbf{b}=0.$ For an ideal fluid $\nu^\prime_k=0$ the
equation \Eqref{upsi=} for $K_y=0$ case takes the form
\begin{equation}
 \mathrm{d}_{\tau}\upsilon_x
 = \alpha(\tau)\upsilon_x
\end{equation}
and have the solution
\begin{equation}
\upsilon_x(\tau)=\frac{1}{1+\tau^2},
\end{equation}
which is depicted at Fig.~1. This time dependence is common for all
$K_z$ for which the dissipation is negligible.
\begin{figure}
\centering \includegraphics[angle=-90,width=8cm]{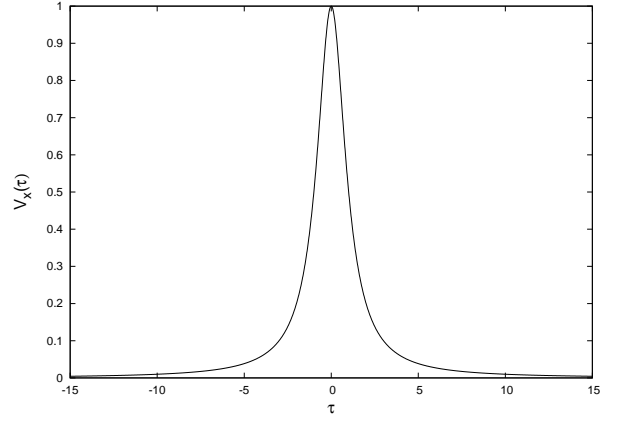} \caption{Time
dependence of velocity in a
shear flow at zero magnetic field; pure hydrodynamic solution. }
\end{figure}

\subsection{Nonzero magnetic field. Analytical solutions}
\label{app:magnetic}

We will perform analysis of the model case for $K_y=0.$ The y-system
\Eqref{upsiy=} and \Eqref{bey=} describes Alfv\'en waves for
which the shear have negligible influence. The shear flow is
important for the slow magnetosonic waves for which the velocity
oscillations and variations of the magnetic field are in the plane
of the wave-vector and the constant external magnetic filed. Only
for this waves we have significant amplification by the shear flow
which we will investigate in the beginning for an ideal fluid for
which the x-system \Eqref{upsi=} and \Eqref{bex=} reads
\begin{eqnarray}
\label{upsix=ap} \mathrm{d}_{\tau}\upsilon_x
 &=& \alpha(\tau)\upsilon_x
 - K_z b_x\\
\label{bex=ap} \mathrm{d}_{\tau} b_x &=& K_z \upsilon_x .
\end{eqnarray}
\begin{figure}
\label{Fig:Omurtag} \centering
\includegraphics[width=8cm]{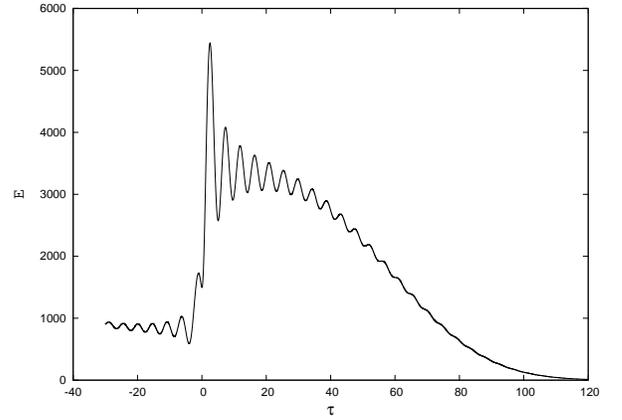} \caption{Energy of an
Alfv\'en wave as function of dimensionless time $\tau.$ An Alfv\'en
wave even at slow viscous damping attenuates heating the plasma and
new one is generated by the turbulence. We the born after billions
years see the light of quasars created by amplification of MHD waves by shear
flow. Parameters of the example: $K_z=0.7,$ $\nu'_m=0.00001,$
$\nu'_k=0.00001,$ $\tau \in (-30, 120),$ $b(-30)=0,$ and
$v(-30)=1.$}
\end{figure}
\begin{figure}
\label{Fig:zero_damping} \centering
\includegraphics[width=8cm]{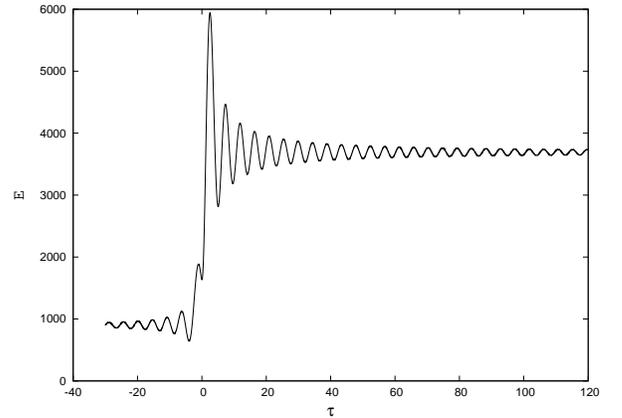} \caption{``Lasing of
alfvons'' as at zero damping. Amplification of energy of the slow
magnetosonic wave by shear flow. Parameters: the same as
Fig. 2 but without friction $\nu'_m=\nu'_k=0.$}
\end{figure}
The substitution of $\mathrm{d}_{\tau}\upsilon_x$ from
\Eqref{upsix=ap} in time differentiated of \Eqref{bex=ap}
gives
\begin{equation}
\mathrm{d}_{\tau}^2 b_x + \alpha(\tau) \mathrm{d}_{\tau} b_x +
K_z^2b_x=0.
\end{equation}
With the help of the substitution
\begin{equation}
b_x(\tau)=\psi(\tau)\sqrt{1+\tau^2}
\end{equation}
we arrive at the effective Schr\"odinger equation for the alfvon
amplitude
\begin{equation}
\label{Schroedinger}
\mathrm{d}_{\tau}^2
\psi+\left(K_z^2-\frac{1}{(1+\tau^2)^2}\right)\psi
=\mathrm{d}_{\tau}^2
\psi+\frac{2m}{\hbar^2}
\left(\tilde E-\tilde U\right)\psi=0.
\end{equation}
\begin{figure}
\label{Fig:psi_amplification} \centering
\includegraphics[width=8cm]{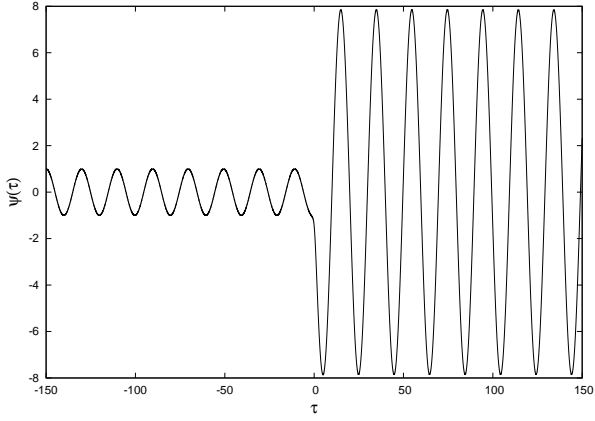} \caption{``Lasing of
alfvons'': the time dependence of the fictitious Schr\"odinger
equation amplitude $\psi(\tau)\propto B_x(\tau)/\sqrt{1+\tau^2}$
for: $K_z=\sqrt{0.1},$ $\psi(-150)=1,$
$\mathrm{d}_\tau\psi(-150)=0.$}
\end{figure}
\begin{figure}
\label{Fig:potential}
\centering \includegraphics[angle=-90,width=8cm]{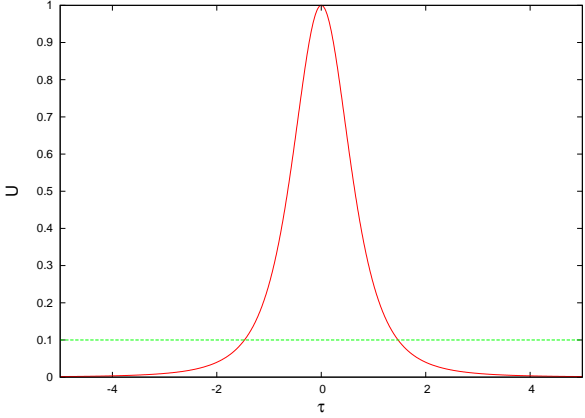}
\caption{Effective potential $1/(1+\tau^2)^2$
and effective energy $K_z^2=0.1$ for the solution of Schr\"odinger equation depicted at
Fig. 4 .}
\end{figure}
For small enough effective energies $K_z^2$ we have a classical
forbidden region where the magnetosonic waves are amplified.

This equation has two linearly independent solutions even (g) and
odd (u)
\begin{eqnarray}
\psi_\mathrm{g}(0)&=&1,\quad \mathrm{d}_\tau\psi_\mathrm{g}(0)=0,\\
\psi_\mathrm{u}(0)&=&0,\quad \mathrm{d}_\tau\psi_\mathrm{u}(0)=1,
\end{eqnarray}
\begin{figure}
\centering
\includegraphics[width=8cm]{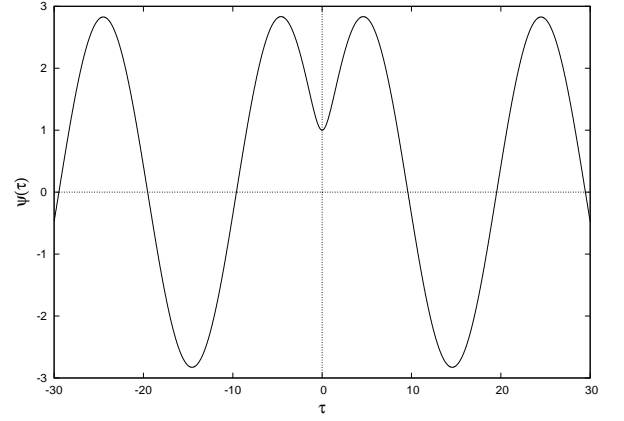}
\caption{Even wave eigenfunction: $\psi_\mathrm{g}(0)=1,$
$\mathrm{d}_\tau\psi_\mathrm{g}(0)=0.$}
\end{figure}
\begin{figure}
\centering
\includegraphics[width=8cm]{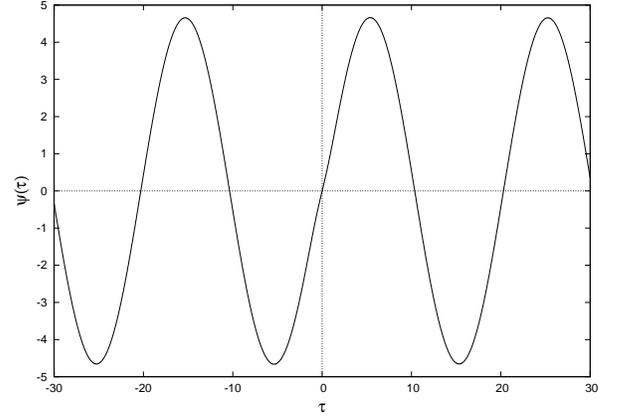}
\caption{Odd wave eigenfunction: $\psi_\mathrm{u}(0)=0,$
$\mathrm{d}_\tau\psi_\mathrm{u}(0)=1$.}
\end{figure}
\begin{eqnarray}
\psi_\mathrm{g}\!=\sqrt{1+\tau^2}\mathrm{H_c}
(0,-\frac{1}{2},0,-\frac{K_z^2}{4}, \frac{1+K_z^2}{4};-\tau^2),\\
\psi_\mathrm{u}\!=\tau \sqrt{1+\tau^2}\mathrm{H_c}
(0,+\frac{1}{2},0,-\frac{K_z^2}{4}, \frac{1+K_z^2}{4};-\tau^2),
\end{eqnarray}
and the general solution we will represent by the linear combination
\be
\psi(\tau)=C_\mathrm{g}\psi_\mathrm{g}(\tau)+C_\mathrm{u}\psi_\mathrm{u}(\tau).
\ee

The confluent Heun function $\mathrm{H_c}(z)$ obeys the differential
equation
\begin{equation}
y''+\left(4p+\frac{\gamma}{z}+\frac{\delta}{z-1}\right)y'
+ \frac{4p\alpha z-\sigma}{z(z-1)} y=0,
\end{equation}
with initial conditions
\begin{equation}
y(0)=0,\qquad y'(0)=\frac{\sigma}{\gamma}
\end{equation}
and close to $z=0$ has the Taylor expansion
\begin{equation}
\mathrm{H_c}(p,\alpha,\gamma,\delta,\sigma;z)=\sum_{n=0}^\infty C_n
z^n,
\end{equation}
where for the coefficients we have the initial conditions $C_{-1}=0$
and $C_0=1$ and recursion
\begin{equation}
C_{n+1}=-(g_n C_{n}+ h_n C_{n-1})/f_n, \quad n=0,1,2,\dots
\end{equation}
where
\begin{eqnarray}
g_n&=&n(n-4p+\gamma+\delta-1)-\sigma,\\
h_n&=&4p(n+\alpha-1),\\
f_n&=&-(n+1)(n+\gamma).
\end{eqnarray}

For large enough arguments we have the expansion
\begin{equation}
\mathrm{H_c}(p,\alpha,\gamma,\delta,\sigma;z)=\sum_{n=0}^\infty C_n^{(\infty)}
z^{\alpha-n},
\end{equation}
where only the recursion functions are different
\begin{eqnarray}
g_n^{(\infty)}&=&(\alpha+n)(n-4p-\gamma-\delta+1)-\sigma,\\
h_n^{(\infty)}&=&-(\alpha+n-1)(\alpha+n-\gamma),\\
f_n^{(\infty)}&=&-4p(n+1).
\end{eqnarray}
%

\subsection{Simple quantum mechanical problem}
In order better to analyze the effective MHD equation
\Eqref{Schroedinger} we will solve the corresponding quantum
mechanical problem when we have tunneling trough the barrier
$2m \tilde U/\hbar^2=1/(1+\tau^2)^2,$ supposing that $\psi$ is a
complex function.
We have falling wave with unit amplitude, reflected wave with
amplitude $R$ and transmitted wave with amplitude $T$
\begin{eqnarray}
\psi(\tau\rightarrow -\infty)&\approx &\exp(iK_z\tau)
  + R\exp(-iK_z\tau),\\
\psi(\tau\rightarrow +\infty)&\approx& T\exp(+iK_z\tau).
\end{eqnarray}
Using the asymptotic of the eigenfunctions
\be
\psi_g\approx \left\{ \begin{array}{l}
D_\mathrm{g}\cos(K_z\tau-\phi_\mathrm{g}),
 \quad\mbox{for  } \tau\rightarrow -\infty,\\
D_\mathrm{g}\cos(K_z\tau+\phi_\mathrm{g}),
 \quad\mbox{for  } \tau\rightarrow +\infty,
\end{array}\right.
\ee
\be
\psi_u\approx \left\{ \begin{array}{l}
 -D_\mathrm{u}\cos(K_z\tau-\phi_\mathrm{u}),
 \quad\mbox{for  } \tau\rightarrow -\infty,\\
\;\;\, D_\mathrm{u}\cos(K_z\tau+\phi_\mathrm{u}),
  \quad\mbox{for  } \tau\rightarrow +\infty
\end{array}\right.
\ee
and solving the matrix problem we obtain
\be
C_\mathrm{g}^\mathrm{(q)}D_\mathrm{g}=\exp(i\phi_\mathrm{g}),\qquad
C_\mathrm{u}^\mathrm{(q)}D_\mathrm{u}=-\exp(i\phi_\mathrm{u})
\ee
Then for the tunneling
coefficient we derive
\be
\mathcal{D}=|T|^2=\mathrm{s_{ug}^2},\qquad
\mathrm{s_{ug}}=\sin(\phi_\mathrm{u}-\phi_\mathrm{g})
\ee
\begin{figure}
\centering \includegraphics[width=8cm]{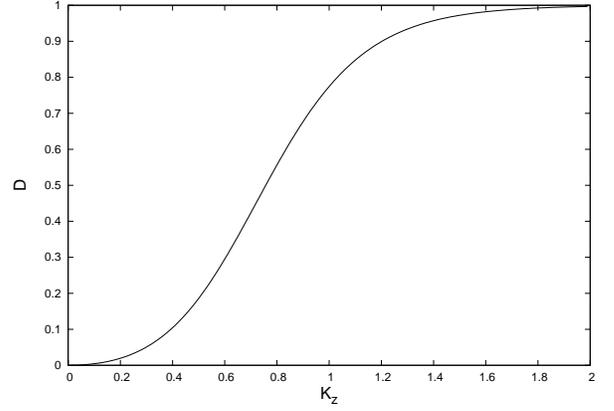}
\caption{Coefficient of transmission through quantum barrier as
function of wave-vector $D=\cos^2\Theta_{ug}$. This coefficient
parameterized amplification of alfvons $A=2/D-1$.}
\end{figure}
Let us introduct notation
\begin{equation}
\varphi_u=\phi_u + \frac{\pi}{2},
\end{equation}
then $\Theta_{gu}=\phi_g - \varphi_u$.

\subsection{MHD and real $\psi$}
For the considered MHD problem $\psi$ is a real variable with
asymptotic
\be \psi \approx \left\{ \begin{array}{l}
\quad \; \cos(K_z\tau-\phi_\mathrm{i}),
 \quad\mbox{for  } \tau\rightarrow -\infty,\\
D_\mathrm{f}\cos(K_z\tau+\phi_\mathrm{f}),
 \quad\mbox{for  } \tau\rightarrow +\infty.
\end{array}\right.
\ee
Solving analogous 2$\times$2 matrix problem we derive
\be
C_\mathrm{g}^{\mathrm{(c)}}D_\mathrm{g}= \mathrm{\frac{s_{iu}}{s_{gu}}}  ,\qquad
C_\mathrm{g}^{\mathrm{(c)}}D_\mathrm{u}= \mathrm{\frac{s_{ig}}{s_{gu}}}
\ee
and for the phase and amplification of the signal we have
\begin{eqnarray}
\tan \phi_\mathrm{f}=\mathrm{\frac{s_{ig} s_{u} + s_{iu} s_{g}}{s_{ig} c_{u} + s_{iu} c_{g}}},\\
\mathcal{A}(\phi_\mathrm{i},K_z^2)=D_\mathrm{f}^2=\frac{\mathcal{N}}{\mathcal{D}},\\
\mathcal{N}= \mathrm{ (s_{ig} s_{u} + s_{iu} s_{g})^2 + (s_{ig} c_{u} + s_{iu} c_{g})^2 },
\end{eqnarray}
where
\begin{eqnarray}
\mathrm{s_{ig}}&=&\sin(\phi_\mathrm{i}-\phi_\mathrm{g}),\quad
\mathrm{s_{iu}}=\sin(\phi_\mathrm{i}-\phi_\mathrm{u}),\\
\mathrm{s_{g}}&=&\sin(\phi_\mathrm{g}),\quad
\mathrm{s_{u}}=\sin(\phi_\mathrm{u}),\\
\mathrm{c_{g}}&=&\cos(\phi_\mathrm{g}),\quad
\mathrm{c_{u}}=\cos(\phi_\mathrm{u}).
\end{eqnarray}
For large enough wave-vectors we have the asymptotic
\begin{equation}
\phi_\mathrm{g}(K_z^2\gg1)=0, \qquad
\phi_\mathrm{u}(K_z^2\gg1)=-\frac{\pi}{2}.
\end{equation}
\begin{figure}
\centering
\includegraphics[width=8cm]{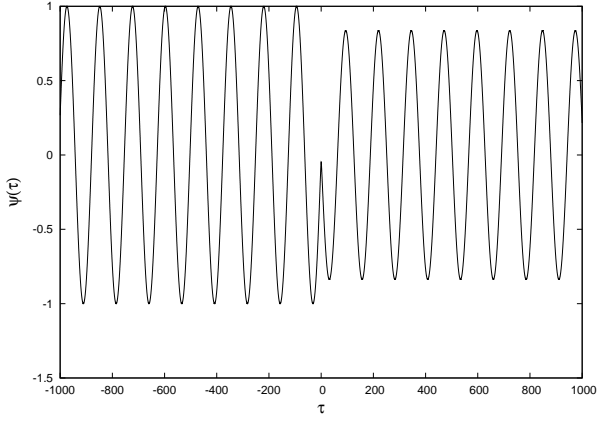}
\caption{Numerical solution  of Schr\"odinger equation for amplitude
of Alfv\'en waves. Property function is approximation by two sinusoid
connected with cusp corresponding of $\delta$-like potential in
relevant quantum mechanical problem. Initial conditions are:
$K_z=0.05$ $\psi(\tau=-1000)=\cos(K_z\tau - \varphi_0)$ ;
$\mathrm{d}_\tau\psi(\tau=-1000)=-K_z\sin(K_z\tau - \varphi_0),$
and $\varphi_0=\frac{\pi}{2}.$}
\end{figure}
\begin{figure}
\centering \includegraphics[width=8cm]{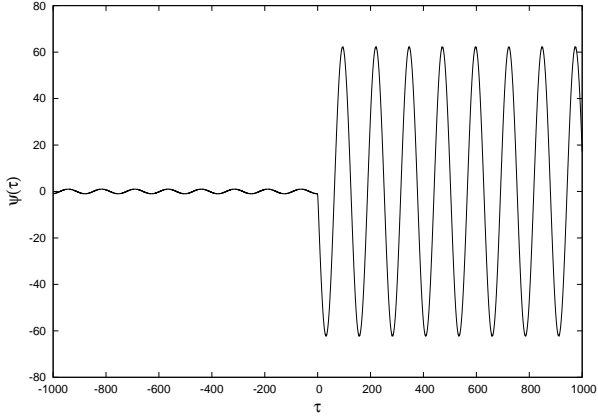}
\caption{Numerical solution of effective Schr\"odinger equation
illustrated amplification of amplitude for small values of
wave-vector. Initial conditions are: $K_z=0.05$  $\psi(\tau=-1000)=\cos(K_z\tau -
\varphi_0)$ ; $\mathrm{d}_\tau\psi(\tau=-1000)=-K_z\sin(K_z\tau - \varphi_0),$ and $\varphi_0=\pi.$}
\end{figure}
\begin{figure}
\centering \includegraphics[width=8cm]{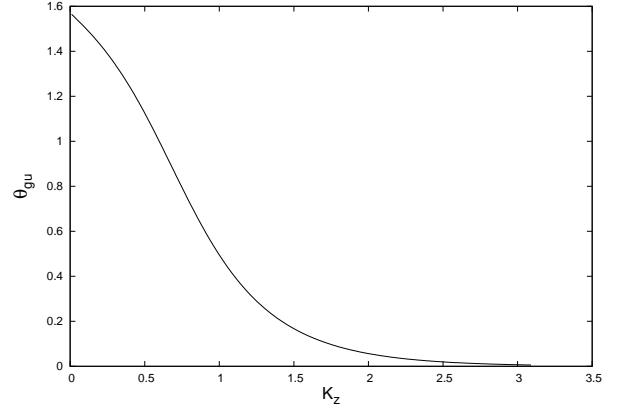}
\caption{Phase analysis of wave equation:the phases of odd $\phi_\mathrm{u}$ and even $\phi_\mathrm{g}$ function of Schr\"odinger equation. Phase difference $\Theta_{gu} = \phi_g - \varphi_u$ parameterized reflection coefficient of quantum mechanical problem and amplification coefficient of MHD problem.}
\end{figure}
\begin{figure}
\centering \includegraphics[width=8cm]{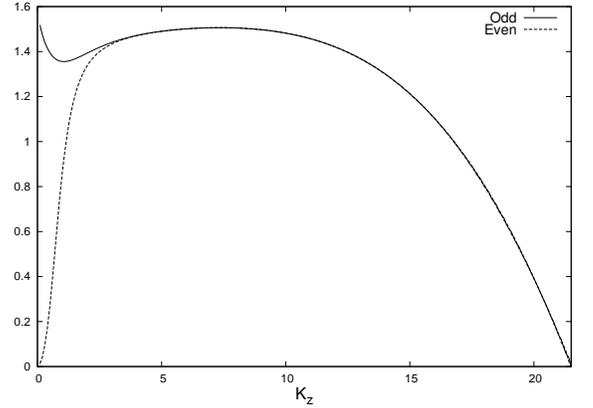}
\caption{Continuous curve above is phase of odd wave function
$\phi_\mathrm{u}$, dots curve under is phase of even wave function
$\phi_\mathrm{g}$ shifted by $\frac{\pi}{2}.$}
\end{figure}

\subsection{Random phase approximation statistical problem}
Finally simple angle averaging
\be
\int_{-\pi}^{\pi}\frac{\mathrm{d}\phi_\mathrm{i}}{\pi}\mathcal{N}(\phi_\mathrm{i})
=2-\mathrm{s_\mathrm{ug}^2}=2-\mathcal{D}
\ee
gives
\be
\label{general_amplification}
\mathcal{A}(K_z^2)=\frac{2}{\mathcal{D}}-1=\frac{2}{\mathrm{s_{ug}^2}}-1.
\ee

In such a way we analyzed the relation between quantum mechanical treatment and MHD one for the effective Schr\"odinger equation.

\subsubsection{Longwavelength approximation}
\begin{figure}
\centering
\includegraphics[angle=-90,width=8cm]{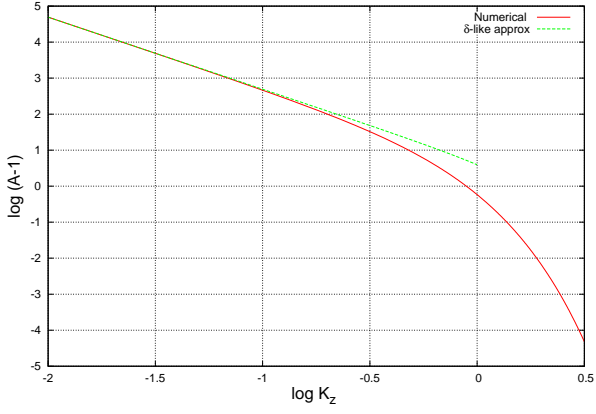}
\caption{Amplification of Alfv\'en waves – keystone of the theory of
heating of accretion disks; amplification in Bell as function of
logarithm of the dimensionless wave-vector; 1 Bell means 10 times
energy amplification. The dotted line is the $\delta$-like long
wavelength approximation which is adequate when the amplification is
significant $\mathcal{A}\gg 1.$  This exact solution at $K_y=0$ is a
test for Monte Carlo calculations in the general case.}
\end{figure}

For small wave-vectors $K_z^2\ll1$ we have $\delta$-potential approximation
in the effective Schr\"odinger equation~Eq.~(\ref{Schroedinger})
\begin{equation}
\frac{1}{(1+\tau^2)^2} \rightarrow \lambda \delta(\tau),\quad
\lambda\equiv\int_{-\infty}^{+\infty}\frac{\mathrm{d}\tau}{(1+\tau^2)^2}=\frac{\pi}{2}.
\end{equation}
The substitution of $\lambda\rightarrow2\lambda$ gives exact\cite{Mishonov:09} result at long-wavelength limit.
The wave function is continuous at $\tau=0$ i.e.
\begin{equation}
\label{zero}
\psi(-0)=\psi(+0),
\end{equation}
but the first derivative have a jump which can be calculated integrating Eq.~(\ref{Schroedinger})
in a small vicinity of $\tau=0$
\begin{equation}
\label{zero_prime}
\mathrm{d}_{\tau}\psi(+0)-\mathrm{d}_{\tau}\psi(-0)=\lambda \psi(0),
\end{equation}
we use obvious alleviation of the notations.
For the wave function
\begin{equation}
\psi(\tau<0)= \cos(K_z\tau-\phi_i), \quad
\psi(\tau>0)= D_f\cos(K_z\tau+\phi_f)
\end{equation}
one can easily solve the equations Eq.~(\ref{zero}) and Eq.~(\ref{zero_prime})
which gives
\begin{eqnarray}
D_\mathrm{f}=\sqrt{1+\frac{\lambda}{K_z} \cos(2\phi)+\frac{\lambda^2}{K_z^2}\cos^2(\phi)},
\\ \phi_\mathrm{f}=\arccos(\frac{\cos\phi_\mathrm{i}}{D_\mathrm{f}}).
\end{eqnarray}
The averaging of the amplification coefficient with respect of the initial phase gives
\begin{equation}
\label{long_amplification}
\mathcal{A}=\int_0^\pi D_\mathrm{f}^2\frac{\mathrm{d}\phi_\mathrm{i}}{\pi}=
1+\frac{\lambda^2}{2K_z^2}\approx \frac{\pi^2}{8K_z^2}=
\frac{1}{8} \left(\frac{\pi A}{k_z V_A}\right)^2\gg 1.
\end{equation}
In other words the Alfv\'en waves amplification can be significant for the long waves.
For the eigenfunctions in this approximation we have
\begin{eqnarray}
\psi_\mathrm{g}&\approx&-\frac{1}{\gamma_\mathrm{g}}\sin(K_z|\tau|-\gamma_\mathrm{g}),\quad K_z>0,\\ \gamma_\mathrm{g}&=&\frac{2K_z}{\lambda}=\frac{4K_z}{\pi}\ll 1, \\ \phi_\mathrm{g}&=&\frac{\pi}{2}-\gamma_\mathrm{g}=\frac{\pi}{2}-\frac{4K_z}{\pi},\\
\psi_\mathrm{u}&\approx& \frac{1}{K_z}\sin(K_z\tau),\\
\phi_\mathrm{u}&\approx& -\frac{\pi}{2}\quad
\phi_\mathrm{ug}=\gamma_\mathrm{g}-\pi,\\
\mathrm{s_{gu}}&\equiv&\sin(\phi_\mathrm{g}-\phi_\mathrm{u})
 \approx\gamma_\mathrm{g}=\frac{4K_z}{\pi}.
\end{eqnarray}
The substitution of this approximate formula for $\mathrm{s_{gu}}$ in the general formula for the amplification Eq.~(\ref{general_amplification}) reproduces the long wavelength result Eq.~(\ref{long_amplification}).

\end{appendix}

\end{document}